\documentclass[a4paper,10pt]{article}
\pdfoutput=1 

\usepackage{jheppub} 
\usepackage[export]{adjustbox}
\usepackage[utf8]{inputenc}
\usepackage{multirow}
\usepackage{bbold}
\usepackage[table]{xcolor}
\usepackage{slashed}
\usepackage{bm}
\usepackage{subfig}
\usepackage{float}
\usepackage{fancyvrb}
\usepackage{fvextra}
\usepackage{soul}
\usepackage{comment}
\usepackage[title]{appendix}

\usepackage{subfig}
\usepackage{color,graphicx,slashed,hyperref}
\usepackage[utf8]{inputenc}
\usepackage{comment}
\usepackage{cancel} 
\usepackage{csquotes} 
 \usepackage{verbatim}
\usepackage{setspace}

\usepackage{booktabs}
\usepackage{amstext} 
\usepackage{array}   
\newcolumntype{C}{>{$}c<{$}} 
\newcolumntype{L}{>{$}l<{$}} 

\newcommand{\sux}{SU(2)_f}
\newcommand{\dralgo}{\texttt{DRalgo}}

\title{Gravitational waves from flavoured SU(2) early-universe phase transitions}

\author[a]{Anna Chrysostomou,}
\emailAdd{chrysostomou@lpthe.jussieu.fr}
\author[b,c]{Alan S. Cornell,}
\emailAdd{acornell@uj.ac.za}
\author[d]{Luc Darm\'e,}
\emailAdd{l.darme@ip2i.in2p3.fr}
\author[d,b]{Aldo Deandrea,}
\emailAdd{deandrea@ip2i.in2p3.fr}
\author[e,f]{Thibault Demartini}
\emailAdd{thibault.demartini@cea.fr}

\affiliation[a]{Laboratoire de Physique Th\'eorique et Hautes \'Energies - LPTHE, Sorbonne Universit\'e, CNRS, 4 Place Jussieu, 75005 Paris, France}
\affiliation[b]{Department of Physics, University of Johannesburg, 
PO Box 524, Auckland Park 2006, South Africa}
\affiliation[c]{Physics Department, De La Salle University, 2401 Taft Avenue, Manila, 1004 Philippines}
\affiliation[d]{Universit\'e Claude Bernard Lyon 1, 
Institut de Physique des 2 Infinis de Lyon, \\ 
CNRS/IN2P3, UMR 5822, F-69622, Villeurbanne, France}
\affiliation[e]{CEA, DAM, DIF, 91297 Arpajon, France}
\affiliation[f]{Université Paris-Saclay, CEA, Laboratoire Matière en Conditions Extrêmes, 91680 Bruyères-Le-Châtel, France}

\abstract{Flavourful extensions of the Standard Model aimed at explaining its fermionic mass structure typically rely on symmetries, broken at high-energy scales far beyond the reach of foreseeable direct collider searches.
We illustrate, using a $SU(2)$ flavour gauge group, that the breaking of these symmetries up to scales as high as $10^7$ GeV could generate a gravitational wave signal potentially observable by future observatories.  
We use dimensional reduction techniques to obtain the finite-temperature effective potential and study the possible first-order phase transitions. We match these transitions to steady-state hydrodynamical solutions in order to determine the corresponding gravitational-wave spectra. We observe that order-one gauge couplings are always required for a first-order phase transition to occur. On the other hand, adding leptoquarks (as an example of particles that are typically present in a complete flavour theory) significantly extends the testable parameter space. We find excellent prospects at the Einstein Telescope for future gravitational-wave detection of flavoured $SU(2)$ early-universe phase
transitions.}


\begin{document} 
\maketitle

\section{Introduction}

The Standard Model (SM) of particle physics constitutes the most accurate and validated framework for describing fundamental interactions. Nevertheless, it exhibits several unsatisfactory features, such as the unexplained hierarchical patterns observed in the fermion masses. This so-called \enquote{flavour-problem} triggered an extensive literature dating back several decades, aimed at naturally explaining this pattern (see for example Ref.~\cite{Feruglio:2015jfa}). 
Starting from the observation that the SM has a $U(3)^5$ global symmetry in the absence of Yukawa interactions, the key idea underlying most of the existing flavour model building is the introduction of new flavour symmetries whose breaking, parameterised by small spurions, controls the shape of the SM Yukawa matrices and, from there, the SM fermions' mass spectrum. Such theories have been suggested since the late $20^{\rm{th}}$-century with various groups structures~\cite{Barbieri:1996ww,Berezhiani:1983hm,Berezhiani:1985in}, down to the $U(1)$ Abelian group leading to the so-called Frogatt-Nielsen mechanism~\cite{Froggatt:1978nt}.

Among these setups, constructions based on gauged flavour symmetries are particularly interesting. They naturally explain the presence of small spurions through dynamical spontaneous symmetry breaking (SSB) of the flavour symmetries, and the case of $SU(2)$ flavour gauge groups saw a renewal of interest in recent years~\cite{Chiang:2017vcl,Guadagnoli:2018ojc,Belfatto:2018cfo,Belfatto:2019swo,Carvunis:2020exc,Darme:2022uzl,Antusch:2023shi,Greljo:2024zrj,Altmannshofer:2024hmr,Mohanta:2024wcr}. 
Despite the diversity of model constructions, several common features can always be identified: (1) the new flavour group -- either global or local -- \textit{must} be broken, with a Higgs-like mechanism often favoured in the second case; (2) the presence of additional degrees of freedom (such as new vector-like fermions (VLFs), scalar leptoquarks (LQs), etc.) is ubiquitous. The presence of flavour symmetries typically prevents the appearance of many SM Yukawa couplings prior to their breaking. They must be generated a posteriori, either via loops or at the effective-level from, for example, five-dimensional operators including flavour spurions which then require additional particles in the full theory.

In many cases, the extreme precision of experimental searches for very rare meson or lepton decay processes suggests -- in the absence of model-dependent coupling suppressions -- an extremely large flavour scale at which the SM flavour patterns would be generated. 
For instance, unsuppressed neutral flavour-changing interactions among first and second fermionic generations are experimentally constrained to energy scales up to $\mathcal{O}(1000) $ TeV~\cite{Feruglio:2015jfa}. At such energies, well above those currently accessible to colliders, gravitational wave (GW) detectors \cite{WhitePaper2022_GWphysics,Caprini:2018mtu,Athron2023_GWreview} may allow for new search opportunities. 

Indeed, one may wonder if the dynamical generation of the flavour-breaking spurion should be accompanied by a cosmological phase transition, generating a stochastic GW background (SGWB); such a GW signal may be observable if the flavour symmetry breaking is generated dynamically by a strong first-order phase transition (FOPT) \cite{Witten:1984rs,MazumdarWhite2018_PTs,Caprini:2019_GW-PT-LISA}. This FOPT would mark a dramatic shift in the state of the universe as the latter cools, with the metastable \enquote{false} vacuum (unbroken/symmetric phase) decaying into the \enquote{true} stable vacuum (broken phase) \cite{Coleman:1977py_FateFalseVacI,Callan:1977pt_FateFalseVacII}. This process occurs via the nucleation, expansion, and collision of true vacuum bubbles, ultimately converting the universe from one phase to another \cite{Enqvist:1991xw,Ignatius:1993qn,Espinosa:2010_EnergyBudget}. A critical requirement for a strong FOPT is the presence of many new degrees of freedom whose masses are directly impacted by the symmetry-breaking vacuum expectation value of the scalar field undergoing the phase transition~\cite{Espinosa:2010_EnergyBudget}. Flavour models thus seem to be particularly good candidates in this respect (see for example the recent work by Ref.~\cite{Fabri:2025fsc}). 
Additionally, present flavour constraints would put this model in the frequency range of proposed GW observatories, such as the third-generation ground-based observatory, the Einstein telescope~\cite{Punturo:2010zz} (and possibly the BBO project~\cite{Harry:2006fi}).

\par In this work, to provide a quantitative proof-of-principle, we choose to explore the extension of the SM flavour sector through a new horizontal $SU(2)_f$ flavour gauge group, in which light generations of left-handed fermions transform as doublets \cite{DarmeDeandrea2023_SU2,Greljo:2024zrj}. $SU(2)_f$ is an important ingredient in larger flavour constructions with a different phenomenology from that of an Abelian flavour gauge group. For such a framework, the model constraints are primarily dictated by specific \enquote{flavour-transfer} operators rather than generic flavour-changing currents. The new gauge symmetry breaks well above the electroweak scale, with the associated phase transition arising from the spontaneous breaking of this $SU(2)_{f}$ symmetry by a real SM-singlet scalar $\Phi$, which transforms as a doublet under $\sux$. Our goal is to determine the parameter space where the phase transition is first-order and strong enough to generate a detectable GW signal. Although toy $SU(2)$ extensions of the SM have been studied in the context of phase transitions, particularly within \enquote{dark Higgs} scenarios \cite{EkstedtSchichoTenkanen:2024etx,Ghosh:2020ipy,Badger:2024ekb}, we focus here on the implication of horizontal $SU(2)_f$ flavour models which: (1) include a direct interaction with the SM fields which impact the effective potential; (2) include additional degrees of freedom around the phase transition energy scale in the form of new colour-charged LQs. The latter addition is motivated in part by the explicit realisation of the SM flavour-patterns from Ref.~\cite{Greljo:2024zrj}, and mostly for the need to illustrate that the presence of these additional degrees of freedom, ubiquitous in complete flavour theories, can significantly alter the phase transitions. 

In order to obtain the GW spectrum for each parameter point, we leverage in part existing code while developing our own routines for specific points. The thermal effective potential is obtained using dimensional reduction (DR) techniques as given by the \texttt{DRalgo} code~\cite{DRalgo}, although we additionally check the agreement with standard resummation techniques. In finding  the nucleation rates, we further use the \texttt{FindBounce} numerical routines~\cite{FindBounce} to obtain the thermal bounce solution for our effective potential. Once a nucleation temperature and the main parameters of the phase transitions are obtained, we rely on the effective approach in Ref.~\cite{Espinosa:2010_EnergyBudget} thereby introducing an overall friction parameter to describe the out-of-equilibrium effects, which are in our case driven (in large part) by the flavour gauge bosons. We then numerically solve the hydrodynamics to find the possible steady-state solutions for the phase transition wall, which in turn gives us the required parameters to input into global fitted spectra from large-scale simulations~\cite{Caprini:2019_GW-PT-LISA}. The projected experimental reach for the Einstein Telescope~\cite{Punturo:2010zz} and BBO~\cite{Harry:2006fi} is finally based on a simple Signal-to-Noise-Ratio (SNR) approach as is customary in the literature.   

The remainder of this work is structured as follows: in Section \ref{sec:su2}, we describe the $SU(2)_f$ flavour model and the current flavour and collider constraints. In Section \ref{sec:firstorder}, we describe the derivation of the effective potential and the implications for the presence and strength of a FOPT. In Section \ref{sec:gw}, we discuss first the hydrodynamical aspects of the FOPT, and secondly the GW signatures of $SU(2)_f$ models at future experiments.  We finally give our conclusions in Section \ref{sec:conclusion}.

\section{$\sux$ flavour-gauge constructions}
\label{sec:su2}

Gauging the flavour symmetries allows one to ground firmly the Yukawa matrices' structure and provide a useful mechanism to generate the spurions from SSB.
In this section, we briefly review the flavour-physics models relying on an additional $\sux$ flavour symmetry that we will consider in this work. 

\subsection{From simplified $\sux$ to more complete models of flavour \label{subsec:model}}

The starting point of our flavour models is a $SU(2)$ symmetry structure. This symmetry is a key ingredient in a variety of solutions to the flavour problem, as it allows us to distinguish the first two generations from the third. In the absence of new chiral fermions, the possible anomaly-free choices are limited and classified by SM fermions combined in $\sux$ doublet or triplet representations according to the following classification~\cite{DarmeDeandrea2023_SU2}:
\begin{itemize}
    \item (LH): with $Q_L, \ell_L$ in $\sux$ multiplets, or  (RH): with $u_R, d_R,  e_R$ in $\sux$ multiplets.
    \item (B): with $u_R, d_R,  Q_L$ in $\sux$ multiplets or (L): with $\ell_L,   e_R$ in $\sux$ multiplets.
      \item (M1): with $u_R, Q_L,  e_R$ in $\sux$ multiplets or (M2): with $d_R, \ell_L$ in $\sux$ multiplets.
\end{itemize}
Each pair of choices can be gauged simultaneously.

Breaking of $\sux$ can occur through a variety of scalar structures; here, we will focus on a new scalar doublet $\Phi$ in the fundamental representation of $\sux$. This determines to a large extent the effective potential structure, with the actual SM fermions in $\sux$ representations having an impact mostly at NNLO, as we show in the next sections.

This structure is not enough to generate the complete mass matrices for the SM fermions. Since the $\sux$ gauge group does not distinguish between two of the fermionic generations (in the example above, the first and second), all $\sux$ spurions $Y_i$ are \textit{a priori} accompanied by the dual spurion $\epsilon_{ij} Y^{\dagger ,j}$, effectively preventing the mass hierarchy between the first and second generations. This well-known issue can be circumvented by either enlarging the symmetry group -- for instance, including an extra $U(1)$ group -- or by ensuring that each spurion contributes only to the mass of a single generation. In order to obtain a spurion from the $\sux$ SSB, one typically introduces extra fermionic (spurions typically obtained at tree-level) or bosonic (spurions generated at loop-level) states. 

In order to estimate the effects of these auxiliary fields in a self-contained model, we will consider as a practical example in the rest of this work the \enquote{rank-rising} construction from Ref.~\cite{Greljo:2023bix}. Based on the $\sux$ (L) scenario, it further includes VLF-generating second-generation fermionic masses and LQ scalars, tasked to generate the first-generation masses at loop-level. In addition to the left-handed SM fermions associated to $\sux$ doublets and singlet -- \textit{viz.} $q_L^\alpha, q^3_{L}, \ell^\alpha_L$ and $\ell^3_L$ -- the right-handed fermions are assigned to $\sux$ singlets. The additional states are shown in Table~\ref{tab:model}. Given the typical range of flavour constraints, the symmetry-breaking scale $v_{\phi}$ is expected to be in the hundreds of TeV range, with masses of VLFs scaling as $m_{\rm VLF} \gg v_{\phi} \gg v_{\rm EW}$, where $v_{\phi}$ is the vacuum expectation value (VEV) for $\Phi$ and $v_{\rm EW}$ is the electroweak VEV.
\begin{table}[t]
    \centering
    \setlength\tabcolsep{0.1cm}
    \def\arraystretch{1.5}
    \begin{tabular}{@{}CCCCCC@{}}
    \hline\noalign{\smallskip}
       Field & SU(3)_C  & SU(2)_L & U(1)_Y & SU(2)_{\ell + q} & DoF \\
       \hline
    Q_{L,R} & 3 & 2 & 1/6 & 1 & 3 \times 2 \times 2 \times 2 = 24  \\ 
   L_{L,R} & 1 & 2 & -1/2 & 1 & 2 \times 2 \times 2 = 8 \\
    \hline
     \Phi & 1 & 1 & 0 & 2 &1 \\
     S & 3 & 1 & 2/3 & 2 & 3 \times 2 \times 2 = 12 \\
     R_u & 3 & 2 & 7/6 & 1 & 3 \times 2 \times 2 = 12 \\
R_d & 3 & 2 & 1/6 & 1 & 3 \times 2 \times 2 = 12 \\
 \hline\noalign{\smallskip}
    \end{tabular}
        \caption{\textit{Example of new particle content of the full flavour theory from Ref.~\cite{Greljo:2023bix}, $\Phi$ breaks $SU(2)_{\ell+q}$ symmetry to produce Yukawa matrices of rank 1; VLFs $Q_{L,R}$ and $L_{L,R}$ lift the rank of Yukawa matrices to 2 (mass to gen-2); LQs contribute to radiative mass generation in gen-1, lifting Yukawa matrices to rank 3. The two last LQ $R_u$ and $R_d$ will be assumed massive and are not included in our model.}}
    \label{tab:model}
\end{table}
While this is only one particular example, we expect on general grounds: (1) complete models of flavour, based on additional gauge symmetries, will be accompanied by new scalar or fermionic fields, generating the spurions and creating
the Yukawa couplings hierarchy; (2) these fields cannot be completely decoupled from the spectrum, although they may be several orders of magnitude heavier than the SSB scale. 

In order to assess the importance of these extra degrees of freedom in the dynamics of the phase transition and in the corresponding GW signatures, we consider separately two cases:
\begin{itemize}
    \item the $\sux$-only case, in which only the SSB scalar field and the flavour gauge bosons are assumed to contribute (thus setting $m_{S}, m_{ R_u}, m_{ R_d}~\gg~v_{\phi}$). 
    \item the $\sux + LQ$ case, in which we further include the $S$ LQ (thus setting $m_{ R_u}, m_{ R_d} ~\gg~m_{S}, v_{\phi}$). 
\end{itemize}
Both cases are viable from the point of view of the flavour problem, since the LQs only radiatively contribute to the spurions~\cite{Greljo:2023bix}.
Note that heavy fields, with masses $m(v_{\phi}) \gg T_c$, decouple from the thermal plasma and therefore do not contribute to the degrees of freedom in the broken phase. For the purposes of this work, we will neglect their contributions to the effective relativistic degrees of freedom.

The most important part of our model building is the scalar potential of this theory. Given that the $S$ and $\Phi$ fields have completely distinct quantum numbers, only hermitian mixing quartic is allowed, leading to:
\begin{align}
    V(\Phi, S) \supset  -\mu^2_{\phi} \Phi^{\dagger} \Phi  +\lambda_{\phi} (\Phi^{\dagger} \Phi )^2  +  \mu^2_s S^{\dagger} S  + \lambda_s ( S^{\dagger} S )^2 + \lambda_{\phi s} (\Phi^{\dagger} \Phi) ( S^{\dagger} S ) \;.
\end{align}
Regarding the vacuum structure of the theory, it is critical that the LQ field does not develop a VEV, to prevent the appearance of a colour-breaking vacuum. The co-positivity of the quartic matrices leads to the usual constraints:
\begin{align}
\label{eq:vacstab1}
    \lambda_s > 0 , \; \lambda_\phi >0 , \; \lambda_{\phi s} > - 2 \sqrt{\lambda_\phi \lambda_s} \ ,
\end{align}
to prevent unbounded direction in the potential. Next, in order to prevent the appearance of this second vacuum with non-zero  VEV for the $S$ field, we impose: 

\begin{align}
\label{eq:vacstab2}
    \lambda_{\phi s} > -2 \lambda_s \;\frac{m_\phi^2}{m_s^2} \;,\nonumber \\
        \lambda_{\phi s} > -2 \lambda_\phi \;\frac{m_s^2}{m_\phi^2} \;,
\end{align}
which translates the requirement that a very light LQ with large negative quartic mixing will naturally have direction in the vacuum space with an overall negative squared mass term and develop a VEV of their own.  Although the exact criterion for (meta)-stability would require a careful study of this new vacuum (and, in particular, of its global or local nature), we restrict ourselves to the constraints from Eqs.~ (\ref{eq:vacstab1}) and~ (\ref{eq:vacstab2}). Indeed, in the presence of a second vacuum, our estimate of the tunnelling rate will likely become unreliable due to the possible two-step processes.

\subsection{Flavour transfers and constraints}

\par When the scalar field $\Phi$ acquires the VEV $v_{\phi}$, breaking the horizontal $\sux$ flavour symmetry, the \enquote{W-like} flavour gauge bosons become massive. This symmetry breaking then sets the scale at which flavour-violating transitions occur. Furthermore, these gauge bosons facilitate \enquote{flavour transfer}, where flavour-violating transitions are linked across sectors of the extended SM. That is, a flavour-violating transition ($\Delta F_f$) in one fermionic sector will be pairwise related to a flavour-violating transition ($\Delta F_f^{\prime}$) in another, such that four-fermion operators originating from these flavour gauge boson exchanges always satisfy a null sum rule $\Delta F_f + \Delta F_f^{\prime} = 0$. In other words, these gauge bosons facilitate flavour-changing processes between different fermion sectors. Through these interactions, the gauge bosons create four-fermion operators, linking flavour transitions between quarks and leptons while satisfying anomaly cancellation. Moreover, the large VEV of $\Phi$ introduces suppression factors for these transitions, such that flavour-violating processes are highly constrained and suppressed in the low-energy regime. 

As noted in Ref.~\cite{DarmeDeandrea2023_SU2}, even in the absence of spurions, these flavour-transfer operators can generate very strong constraints on $v_{\phi}$. 
With the flavour gauge symmetry chosen as an example, the SM left-handed fermions are combined in a doublet $Q_{L,i}, L_{i}$ and a singlet of $SU(2)_f$. While a natural choice is to choose the doublet to combine the first and second generation (a choice we will refer to as the $(12)_{f}$ scenario), other flavour alignments may be possible (such as $(13)_{f}$ and $(23)_{f}$), and the corresponding flavour constraints should be modified.  

In any case, the fact that the new flavour gauge bosons interact with both down quarks and leptons leads to very strong constraints from rare meson decays. Additionally, the necessary presence of the CKM matrix will automatically induce the presence of the pure flavour-violating contribution, either in the down or in the up-quark sector.
Relevant channels depend on the flavour orientation and include, for instance:
\begin{itemize}
    \item   Rare flavour-transferring semi-leptonic meson decays,
    \begin{align}
K \to \pi e \mu \, , B \to K e \mu \, , B \to \pi e \mu \, , 
    \end{align}
    along with the corresponding neutrino processes,
        \begin{align}
K \to \pi \nu \nu \, , \ B \to \pi \nu \nu \, , \ B \to K \nu \nu \ .
    \end{align}
    The latter is particularly important due to the fact that the exact leptonic flavour-alignment does not impact the corresponding constraints. We have for the $1-2$ flavour alignment in the quark sector:
    \begin{align}
    \textrm{BR}(K^+ \to \pi^+ \nu \nu) \sim 1.4 \cdot 10^{-11} \times\left( \frac{100 \, \textrm{TeV}}{v_\phi}\right)^4 \ ,
\end{align}
to be compared with the SM prediction~\cite{DAmbrosio:2022kvb,DAmbrosio:2023irq}:
\begin{align}
    \text{BR}(K^+\to \pi^+\nu \bar{\nu})^{\rm SM}    &= (7.86 \pm 0.61)\times 10^{-11}\,,
\end{align}
which is compatible with the NA62 experimental result:
\begin{align}
 &{\rm Br}(K^+ \to \pi^+ \nu \bar{\nu}) = 10.6^{+4.1}_{-3.5}
\times 10^{-11}\quad  (\textrm{NA62}~\textrm{\cite{NA62:2021zjw}}) \; .
\end{align}
    \item Flavour-transferring fully-leptonic meson decays
        \begin{align}
K \to \pi e \mu \,  \quad B_s \to e \mu  \, , \quad B_d \to e \mu \ . 
    \end{align}
    The Kaonic decay process in particular leads to the strongest constraints on models with $1-2$ flavour alignment, with
\begin{align}
    \textrm{BR}(K_L \to  \mu^\pm e^\pm) = 3.0 \cdot 10^{-11} \left( \frac{200 \, \textrm{TeV}}{v_\phi}\right)^4  \; ,
\end{align}
to be compared with the old but stringent experimental limit,
\begin{align}
    \textrm{BR}(K_L \to  \mu^\pm e^\mp) < 4.7 \times 10^{-12} \qquad (\textrm{BNL}~\textrm{\cite{BNL:1998apv}}) \ .
\end{align}
    \item Muon conversion processes, such as 
    \begin{align}
        \mu \ \textrm{Au} \to e   \ \textrm{Au} \   (\textrm{SINDRUM-II}~\textrm{\cite{SINDRUMII:2006dvw}}) \ .
    \end{align}
    Current constraints from these processes lead to limits of the same order as from $K \to \pi e \mu$ rare decays.
    \item Meson oscillations $D_0 - \bar{D}_0$ , $K_0 - \bar{K}_0$, and the $B_s$ and $B_d$ systems, can only arise -- in the absence of spurions -- from the CKM matrix. As shown in Ref.~\cite{DarmeDeandrea2023_SU2,Greljo:2023bix}, the partial unitarity of the CKM matrix in the $1-2$ or $2-3$ sectors ensures an additional suppression of these constraints. $D_0$-oscillation limits are particularly important for $2-3$ flavour alignment. Given the current fit~\cite{PDG2020} and
\begin{align}
    x_D^{\rm exp} = (4.09 \pm 0.48) \times 10^{-3} \ ,
\end{align}
 noting that there is no definite SM prediction yet, we obtain:
 \begin{align}
     x_D = 5.5 \cdot 10^{-4} \ \times \left( \frac{200 \, \textrm{TeV} }{v_\phi} \right)^4 \ .
 \end{align}
\end{itemize}
In general, given that the precise spurion structures have little influence on the actual dynamics of the phase transition, one can instead leave the flavour structure is left completely free. It was then shown in~\cite{DarmeDeandrea2023_SU2} that a global scan on the flavour parameters led nonetheless to the constraints ${v_\phi} \gtrsim 38$ TeV, which we will retain as the most conservative limit from flavour physics on this class of theories.

Finally, while the LHC searches only extend to the multi-TeV regime at best (see, for instance, Refs.~\cite{CMS:2019buh,CMS:2021ctt,CMS:2022fsw}), they can still probe a viable parameter space with respect to flavour constraints. Indeed, while the latter depends only on $v_{\phi}$, LHC searches typically extend to smaller gauge couplings $g_f \lesssim 0.01$ starting from a few TeV, thus effectively testing values for $v_\phi$ in the hundreds-of-TeV range, as long as the gauge coupling is small enough that the $SU(2)_f$ vector boson is at the TeV scale. Interestingly, we will show that GW searches can probe precisely the \textit{opposite} regime of large gauge couplings and large masses.

\section{Describing the first-order phase transition}
\label{sec:firstorder}

\par A crossover phase transition, such as the SM electroweak phase transition,\footnote{The SM electroweak phase transition has been shown to be a smooth cross-over by analytical calculations~\cite{Kajantie1996_CrossoverSM} and lattice simulations~\cite{DOnofrio2015_SmoothSMcrossover}.} can be promoted to a FOPT by, for example, introducing extra bosonic degrees of freedom in the finite-temperature effective potential. These new degrees of freedom contribute a cubic term to the effective mass, delaying the phase transition and strengthening the barrier between false and true vacua~\cite{Csikor1998_MakeItFOPT}. This establishes a strong FOPT that may be observable at GW detectors. Promising studies suggest that the GW peak frequency for many BSM scenarios falls within the sensitivity range of next-generation detectors (see Ref.~\cite{WhitePaper2022_GWphysics} and references therein). However, precise predictions require incorporation of thermal corrections into the effective potential and accounting for the thermodynamic and hydrodynamic properties of bubble dynamics following the transition.

\par As we shall explain in this section, the resummation of thermal corrections becomes necessary due to the hierarchy of scales that develops at high temperatures between the bosonic zero Matsubara modes ($\omega_n = 0$) and the \enquote{hard} modes with non-zero Matsubara frequencies ($\omega_n \neq 0$), which always carry momenta $p^2 \geq (\pi T)^2$. For light fields with $m \lesssim T$, the zero-mode propagator receives corrections of order $g^2 T^2$ from loops involving hard modes, where $g^2$ denotes a generic quartic coupling. At weak coupling, these loop corrections can become comparable to, or even exceed, the tree-level mass term, thereby invalidating a naive perturbative expansion around the free theory. The standard remedy is to resum these corrections by absorbing the thermal mass shift into a redefined \enquote{tree-level} mass, commonly referred to as the \enquote{thermal mass}. 
Our objective in this section is to demonstrate how to contend with these non-perturbative elements, in order to compute the effective potential and its thermal corrections. We begin with Section \ref{subsec:DRexplained}, introducing DR as a systematic scheme for thermal resummations\footnote{Note that while we review finite-temperature quantum field theory and the resummation techniques in brief, we refer the reader to the reviews, Refs.~\cite{Quiros1998_ICTP,Weir2017_GWsFOPTreview,Croon:2020_ThermalResum,Athron2023_GWreview}, for further details.}, before presenting our implementation of the $\sux$ model in DRalgo in Section \ref{subsec:DRimplementation}.

\subsection{Finite temperature contributions using dimensional reduction \label{subsec:DRexplained}}

\par The calculation of thermal contributions is performed using the \enquote{imaginary-time} formalism proposed by Matsubara \cite{MatsubaraFormalism1955}, and requires a perturbative expansion in the coupling. Within the SM, the expansion parameter is the weak gauge coupling, $g$; in our theory, it is the flavour-gauge coupling, $g_f$. Standard power-counting \cite{Kajantie1995_DRmethod} with respect to the other couplings present in the problem suggests the following ordering:

\begin{equation}
g_f^2 \sim \lambda_\phi \;.
\end{equation}
\noindent With this scaling, we can replace all couplings with appropriate powers of $g_f$ when discussing the issue of power-counting. Furthermore, squared scalar masses scale as $\mu^2 \sim g_f^2 T^2$ and fermionic masses as $\mu^2 \sim g_f T$ \cite{DRalgo}. For infrared bosonic modes of mass $m$ and energy $E \ll T$, this effective loop expansion parameter is enhanced at high temperature, such that 
\begin{equation} 
g_f^2 \rightarrow g^2_{f \; n_{B}} = \frac{g_f^2}{e^{E/T} - 1 } \sim \frac{g_f^2 T}{E} \geq \frac{g_f^2 T}{m} \;.
\end{equation} 
\noindent At sufficiently high temperatures comparable to $m/g_f^2$, the infrared bosonic modes become strongly coupled. 

\par Upon calculating the one-loop correction to the two-point correlator at high temperature with respect to the mass scale, the path integral yields an ultraviolet-divergent zero-temperature piece and an ultraviolet-finite $-$ but infrared-sensitive $-$ temperature-dependent piece. The latter can be separated into \enquote{soft} zero modes that scale as $p\sim g_fT$ and an infinite tower of \enquote{hard} non-zero Matsubara modes that scale as $p \sim \pi T$. The zero mode acquires an effective thermal mass $m^2_T~=~m^2 + \# g_f^2 T^2$ in the high-temperature limit,\footnote{We use the shorthand $\#$ to indicate the coefficient that depends on the group structure and representation of the fields within the BSM framework of interest.} where the thermal mass arises as a screening mass; it is the non-zero mode excitations of the thermal plasma that screen the zero modes. 
As such, we observe an emergent hierarchical separation of energy scales between the \enquote{soft} (\enquote{heavy}, $p \sim g_f T$) and the \enquote{hard} (\enquote{superheavy}, $p \sim \pi T$) scales \cite{DRalgo} (in parentheses, the notation used in Ref.~\cite{Kajantie1995_DRmethod}), respectively, scaling for perturbative coupling $g_f$ as
\begin{equation} \label{eq:scales}
\left( \frac{g_f}{4 \pi} \right) \pi T \ll \pi T \;.
\end{equation}

\noindent However, during the phase transition, the scalar component triggering the flavour breaking has a smaller mass due to the required cancellation between $-\mu_\phi^2<0$ and the dominant thermal contribution $\propto g_f^2 T^2$.
\par We can then construct a soft-scale 3d EFT by integrating out the hard scale (i.e. the $n > 0$ Matsubara modes and the fermions), and then an ultrasoft-scale EFT by integrating out the soft-scale particles (i.e. the temporal scalars with soft-scale Debye masses). The result is an EFT comprised of spatial gauge fields and the light scalar field that drives the phase transition \cite{Croon:2020_ThermalResum}. Energy scales higher than the hard scale are exponentially suppressed (by a Boltzmann factor). Note that the $\pi T$ factors arise from the Matsubara modes while the $(g_f/ 4 \pi)$ factors come from loop integrals. Since the expansion parameter for the hard scale $\epsilon_{\rm hard}~\sim~(g_f/\pi)^2~\ll~1$, the theory is perturbative at this scale; for softer scales, the expansion parameters are larger, indicating slower convergence \cite{Gould2023_Exp}. The ultra-soft EFT can be used down to the so-called ultra-soft scale $\left( g_f/4 \pi \right)^2 \pi T$, where our non-Abelian gauge theory becomes non-perturbative~\cite{Linde1980_TheIRProblem}.\footnote{Intermediate scales denoted by fractional coupling lie between hard and ultra-soft scales. From the ultra-soft scale on the left to the hard scale on the right, these lie on either side of the soft scale, \textit{viz.} the \enquote{super-soft} and the \enquote{semi-soft} scales. In the notation of Refs \cite{Gould2023_Exp,EkstedtSchichoTenkanen:2024etx},
\[
\underset{\rm \underline{ultrasoft}}{\left( \frac{g}{4 \pi} \right)^2 \pi T} \ll \underset{\rm supersoft}{\left( \frac{g}{4 \pi} \right)^{3/2}} \ll \underset{\rm \underline{soft}}{\left( \frac{g}{4 \pi} \right) \pi T} \ll \underset{\rm semisoft}{\left( \frac{g}{4 \pi} \right)^{1/2}} \pi T \ll \underset{\rm \underline{hard}}{\pi T {\color{white}{\frac{1}{1}}}} .
\]
}

Once we recognise this energy scaling and its consequences, the application of an EFT-like DR approach is evidently pragmatic. The underlying objective of the approach is to ensure that perturbative modes are treated perturbatively and
non-perturbative modes are subjected to the appropriate non-perturbative treatment. Fermions and bosonic non-zero Matsubara modes that are integrated out, as well as the bosons of the soft scale, are perturbative; the contribution from modes below the non-perturbative ultra-soft scale can then be treated with non-perturbative lattice approaches \cite{Croon:2020_ThermalResum}. However, as discussed in Ref. \cite{Croon:2020_ThermalResum}, we can still continue to describe the ultra-soft EFT of a weakly-coupled parent theory during the phase transition using two-loop perturbation theory with reasonable accuracy, provided the FOPT is sufficiently strong (as shown in Ref.~\cite{EkstedtSchichoTenkanen:2024etx}, we can expect a precision somewhat below the $\mathcal{O}(1)$-level by using second-order contributions; we will retain this as our theoretical precision). Altogether, there are two key advantages of the EFT approach: (1) since the infrared and ultraviolet degrees of freedom are separated, we avoid double counting; (2) by resumming large logarithmic terms $\log \{ \Lambda_{\rm UV} / \Lambda_{\rm IR} \}$, we avoid hierarchies of scales breaking perturbation theory.

\par With these energy scales in mind, we can proceed with our analysis of the cosmological phase transition for the $\sux$ model. For such an exercise, the first step is the calculation of the effective potential. We discuss the application of the DR approach to our model of interest in Section \ref{subsec:DRimplementation}. Our objective is to compute the 3d effective potential in the \enquote{ultra-soft} theory, for which all heavier modes are decoupled and only the scalar field that triggers the phase transition and the $S$ LQ remains. This is a step-by-step process: we begin at the 4d theory, where we first decouple the towers of thermal $n \neq 0$ modes (i.e. fermions and transverse gauge bosons), integrating out the \enquote{hard scale} to produce the \enquote{soft-scale} 3d effective potential. Then we integrate out the remaining scalars and the temporal (longitudinal) components of the gauge bosons to obtain the \enquote{ultra-soft} effective potential. To relate the 4d to the 3d theory, we use the scaling relations \cite{Croon:2020_ThermalResum},
\begin{align}
    V^{\rm 3d}_{\rm eff}\;  T & \simeq V^{\rm 4d}_{\rm eff} \label{eq:4d3dV} \\
    \phi^{\rm 3d}  \sqrt{T} & \simeq \phi^{\rm 4d} \;. \label{eq:4d3dphi}
\end{align}
\noindent The first expression holds up to the order $\mathcal{O}(g_f^3)$. However, the second only holds at leading order in powers of $\mathcal{O}(g_f^2)$. Beyond this, terms from the momentum-dependent thermal screening contribute to the 3d matching relations and must also be taken into account. 

\par Note that the shape of the effective potential provides key insights into the nature of the phase transition. Bosonic degrees of freedom enhance the effective cubic term in the potential, which manifests as an increased height in the thermal barrier between the degenerate minima at $\phi = 0$ and $\phi = v_{\phi}$  that develop at the critical temperature, $T=T_c$. The sphaleron decoupling condition, $v_{\phi}/T_c \gtrsim 1$, indicates a sufficiently large barrier, a significant separation between the symmetric and broken minima, and thus a discontinuous jump in the phase transition \cite{Dine:1992wr,Quiros1998_ICTP}. Furthermore, the use of a ratio $v_\phi/T_c$ provides a gauge-invariant proxy for identifying strong first-order behaviour in perturbative analyses \cite{Patel2011_GaugeIndependence}. In the next section, we analyse how the behaviour of the model itself influences the phase transition. We shall continue our discussion on how to compute the thermal quantities derived from the 4d effective potential in Section \ref{subsec:PTobs}.

\subsection{Implementation  and loop-stability\label{subsec:DRimplementation}}

\par Let us now consider the application of the DR procedure to the flavour model of interest. The tree-level effective potential of the 4d-theory after symmetry breaking is given by

\begin{align}\label{eqn:V0main}
V^{\rm 4d}_{\textrm{tree}}(\phi)=-\frac{1}{2}m^2_{\phi} \phi^2 +\frac{1}{4}\lambda_{\phi} \phi^4   + V_{\rm LQ} \;,
\end{align}
where we used lower case notation $\phi$ for the lower components of $SU(2)_f$ doublets $\Phi$. The LQ scalar potential $V_{\rm LQ}$ is not included in the pure $SU(2)_f$ case, and is given by 
\begin{align}
 V_{\rm LQ} =  \frac{1}{2} m^2_s s^{\dagger} s + \frac{1}{4} \lambda_s (s^{\dagger} s)^2 + \frac{1}{2} \lambda_{\phi s} \phi^2 (s^{\dagger} s) \;,
\end{align}
so that the real field $\phi$ is weakly-coupled to the complex scalar LQ via the portal coupling $\lambda_{\phi s} \phi^2 \vert s \vert^2/2 $. We detail the construction of the 4d effective potential and the thermal corrections in Appendix \ref{app:4dtheory}. In the rest of this work, we will use the tree-level VEV $v_0$ of the $\phi$ scalar  as our reference energy scale, setting the tree-level negative square mass from $m_\phi^2 = - v_0^2 \lambda_{\phi} $. We further set the initial 4d renormalisation scale to $v_0$ for each of our points.

\par Following the DR techniques described in the last section, at NLO-level, the 3d effective potential is in the ultra-soft limit. In this limit, all degrees of freedom have been decoupled, apart from the spatial component of the gauge fields and the scalar field triggering the phase transition, such that the effective potential is given by:
\begin{align}
    V_{\rm eff}^{\rm US}= \frac{1}{2} (m_\phi^{\text{US}})^2 \phi^2 + \frac{1}{4} \lambda_\phi^{\rm US} \phi^4-\frac{3 \phi^3 (g_f^{\text{US}})^3+12 \left(\phi ^2 \lambda_\phi^{\rm US}+(\mu_\phi^{\text{US}})^2 \right){}^{3/2} + 4 \left(3 \phi ^2 \lambda_\phi^{\rm US}+ (m_\phi^{\text{US}})^2 \right)^{3/2}}{48 \pi } \;.
\end{align}

When considering the case of an additional LQ, we will additionally include them in the ultra-soft EFT, leading to an additional contribution $V_{\rm eff}^{\rm US}$. Indeed, despite acquiring a thermal mass squared proportional to the soft scale $(g T)^2$, group theoretic factors actually make this contribution more than an order of magnitude smaller than the squared Debye mass for the flavour gauge boson that we use in practice as a reference for the soft scale. Thus, as long as their tree-level mass is smaller than $v_0$, they are better described in our low-energy ultra-soft EFT. 
\begin{figure}[t]
\centering
\includegraphics[width=0.7\linewidth]{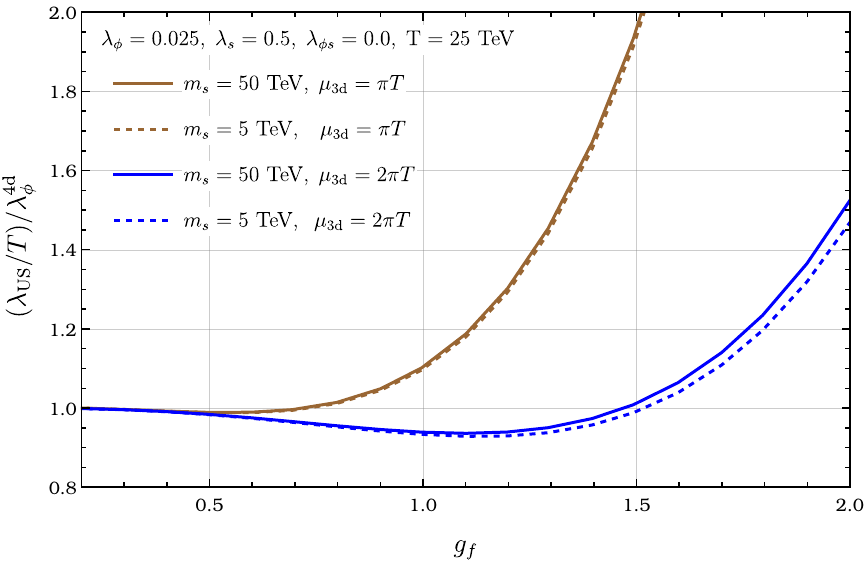}
\caption{Scalar quartic coupling in the ultra-soft theory at the soft matching scale as function of the initial 4d flavour gauge couplings for $\lambda_\phi = 0.025$, for two different 3d to 4d matching scales. We set $v_0 = 50$ TeV.}
\label{fig:3dcoup}
\end{figure}
\begin{figure}[t]
\centering
\includegraphics[width=0.8\linewidth]{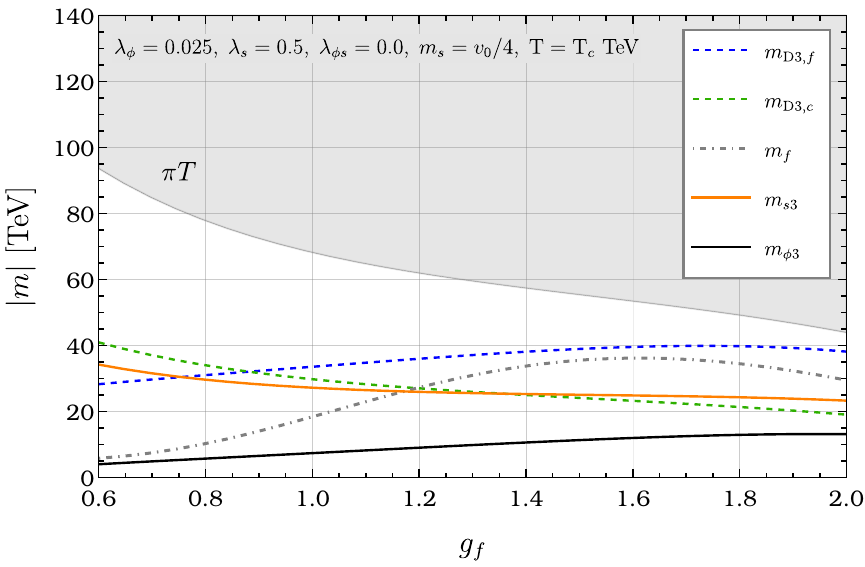}
\caption{\textit{The various mass scales relevant for the 3d soft theories. We set $v_0 = 50$ TeV.}}
\label{fig:3dmasses}
\end{figure}
This is illustrated in Fig.~\ref{fig:3dmasses} where we show the various mass scales in the unbroken vacuum that drives the EFT approach beyond the DR procedure. We observe that in most of the parameter space relevant to our study, we maintain a light scalar $\Phi$, while the temporal scalars along with the LQ $S$ indeed constitute a proper intermediary scale up to $g_f \gtrsim 1.5$. For larger couplings, the Debye mass becomes of the order of the hard scale $\sim \pi \, T$ and one should treat the above process with care~\cite{Kierkla:2023von,Chala:2024xll}. As we show in this section, going to larger gauge couplings does not necessarily lead to a stronger phase transitions in any case due to the $g_f$-induced correction with the scalar quartic $\lambda_\phi$.\\

Based on the cases specified in Section \ref{subsec:model} (i.e. whether the $S$ LQ participates in the phase transition), as well as the constraints of Eqs. (\ref{eq:vacstab1}) and (\ref{eq:vacstab2}), we introduce the following benchmark points: 
\begin{itemize}
    \item BP1 - No LQ mixing,   $\{\lambda_{\phi},  \lambda_{\phi s},  g_f \} = \{0.025, 0.00, 1.5 \}$ ,
    \item BP2 - With very slight LQ mixing,  $\{\lambda_{\phi}, \lambda_{\phi s},  g_f \} = \{0.005,0.01, 1.0 \}$ ,
    \item BP3 - With LQ mixing,   $\{\lambda_{\phi}, \lambda_{\phi s},  g_f \} = \{0.025, 1.0, 1.5 \}$ ,
    \item BP4 - With LQ mixing,   $\{\lambda_{\phi},  \lambda_{\phi s},  g_f \} = \{0.1, 1.5, 1.0 \}$ ,
    \item BP5 - With LQ mixing, $\{\lambda_{\phi},  \lambda_{\phi s},  g_f \} = \{0.005, 0.25, 1.5\}$ ,
\end{itemize}
\noindent with a fixed $\lambda_s = 0.5$. We set the mass of the $S$ LQ as $m_s = v_0 = 50$ TeV, unless otherwise stated; in Section \ref{sec:gw}, we explicitly demonstrate how the GW detection prospects are thereby influenced. We implemented the model within the \texttt{\texttt{DRalgo}} package \cite{DRalgo} which relies upon the \texttt{GroupMath}~\cite{GroupMath} framework to generate the Lagrangian of the theory. All the 4d parameters are defined at the initial scale of $\mu_{\rm ini} = v_0 = 50$ TeV, then evolved to the 3d hard thermal matching scale, chosen be $\pi T$ to keep both bosonic and fermionic $ L_b $ and  $ L_f $ logarithmic contributions of order one. We then integrate out the scalar LQ at the same soft scale as the temporal modes to obtain the ultra-soft theories valid down to the ultra-soft scale, $\mu > g_f^2 T$. The resulting effective 3d potential is calculated at NNLO using \texttt{\texttt{DRalgo}} built-in routines and the 3d parameters. We stress that the matching procedure is realised in the unbroken vacuum and thus assumes that the hard thermal masses are larger than the vacuum-induced one. We discuss this point later in this section.

\vspace{0.6cm}
\par To obtain a strong-enough FOPT, we must rely on a tree-level potential with a relatively small quartic and large gauge coupling. Interestingly, this regime also corresponds to the most out-of-reach part of the parameter space for flavour searches within conventional collider-based experiments, due to the constraints from flavour-transfer observables. The two requirements are, however, incompatible at one and two-loop levels, due to both the large renormalisation group equation (RGE) evolution of the Higgs quartic induced by the $\sux$ gauge bosons and the corresponding matching corrections.
Let us first consider the matching corrections that arise automatically in the DR approach considered in this work.

\par The 3d soft-scale couplings written in terms of their 4d counterparts at NLO. Assuming that $\lambda_\phi \ll g_f^2, \lambda_{\phi s}$ and focusing on the gauge-flavoured section, we can write
\begin{align}
 &\lambda_{\phi}^{\rm 3d} =  T \left[ \lambda_{\phi} + \frac{1}{ (16\pi)^2} \Big(g_f^4\,  (6-9 L_b)+72 g_f^2 \lambda_{\phi} L_b    - 48 L_b
 \left(   \lambda_{\phi \rm{s}} ^2 +4 \lambda_{\phi }^4 \right) \Big) \right]\;, \nonumber \\
    &\left(g_f^{\rm 3d} \right)^2 =  g_f^2 T \left[1+  \frac{g_f^2}{48 \pi ^2}
    \left(20 L_b-15 L_f+2 \right) \right]  \;.
    \label{eq:ScouplingsLQ}
\end{align}
Here we have used the notation from \texttt{DRalgo} for the order-1 logarithmic terms:
\begin{align}
     L_b = 2 \, \gamma  + \log \frac{\mu^2}{(4\pi T)^2} \ ; \quad L_f =  2 \, \gamma  + \log \frac{\mu^2}{(\pi T)^2} \
\end{align}
where both cannot be simultaneously cancelled, but are $\sim \pm 1$ for $\mu \sim \pi T$.
It is already clear at this point that the gauge contribution will overcome the tree-level quartic coupling in the regime $\lambda_\phi \sim 0.01 g_f^4$, thus effectively preventing a scale separation larger than two orders of magnitude between the VEV and the dark scalar mass.

\begin{figure}[t]
\centering
\includegraphics[width=0.49\linewidth]{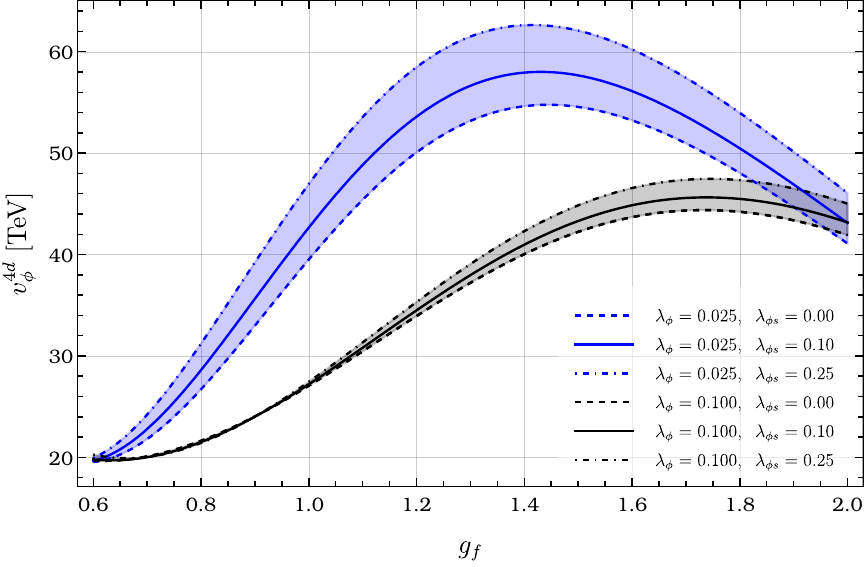}
\includegraphics[width=0.49\linewidth]{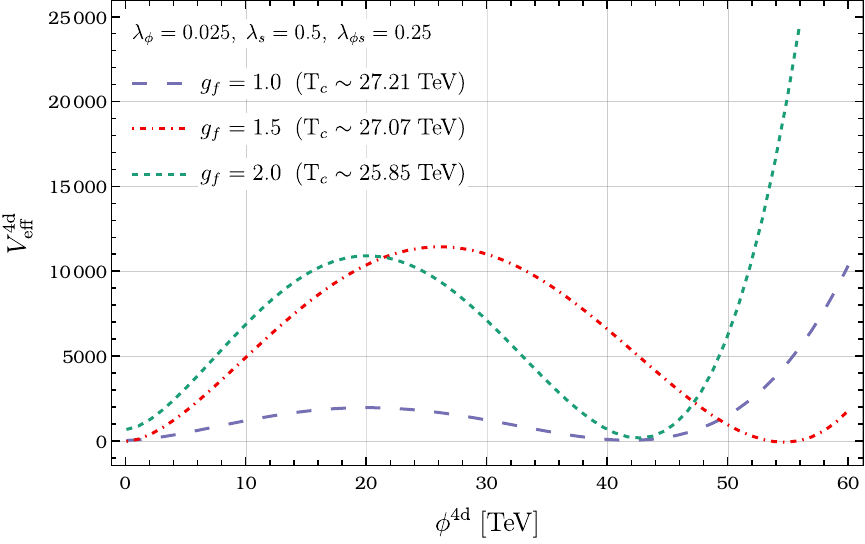}
\caption{\textit{(Left) Evolution of the 4d broken vacuum localisation at the critical temperature from the complete NNLO estimation of \dralgo. (Right)  the 4d effective potential at the critical temperature, with the increasing barrier for $g_f = \{1,1.5,2 \}$, $\lambda_{\phi}=0.025$, $\lambda_s = 0.5$, and $\lambda_{\phi s} = 0.25$. While $T_c$ is equivalent in the 3d and 4d theory, the rescaling between 3d and 4d results in a rescaled VEV in 4d. We set $v_0=50$ TeV. 
}}
\label{fig:VEVingf}
\end{figure}

At the soft scale, we further decouple all the temporal modes. Including only the flavour-gauge part, the explicit \enquote{soft-to-softer} EFT matching relations for the scalar quartic, where the temporal scalar and the singlet are integrated out, reads as 
 
\begin{align}
    &\lambda_\phi^{\rm US} = \lambda_{\phi}^{\rm 3d} - \frac{3}{32 \pi } \left[ \frac{\left(\lambda^{\rm 3d}_{ \phi A_0}\right)^2}{\sqrt{m_{D,f}}}\right]\;, \label{eq:UScouplings}
\end{align}
\noindent where the square brackets contain the quartic mixing $\lambda^{\rm 3d}_{ \phi A_0}$ between the temporal vector and the scalar, with $(\lambda^{\rm 3d}_{ \phi A_0})^2 \sim g_f^2 T/2$ at tree-level and $m_{D,f}$ is the Debye mass. This matching thus tends to reduce the value of the 3d quartic coupling, thereby strengthening the phase transition. Note that when the tree-level LQ mass is larger than the soft scale, we can also simply decouple them together with the temporal scalar. We have checked explicitly that both methods lead to similar results for the thermal parameters presented in the next section.
We show in this case in Fig.~\ref{fig:3dcoup} a combination of both matching on the ultra-soft quartic couplings for different values of the 3d to 4d matching scale. At large couplings the former dominates and the ultra-soft quartic typically increases from its 4d value.
This is a sizeable effect, indeed, in order to understand the effect on the strength of the phase transition, it is instructive to proceed to a simple order of magnitude estimate of the various contributions in the NLO 3d effective potential Eq.~\eqref{eqn:V0main}. As the temperature decreases, the squared mass term becomes gradually smaller, up to the point where the trilinear contributions can lead to a second vacuum with the corresponding barrier. At this point, which is close to the critical temperature, the second vacuum is located at
\begin{align}
v_{\phi} \equiv  \langle \phi\rangle_{T_c} \sim  \frac{\sqrt{(3\kappa_3)^2-4 \lambda_\phi^{\rm US} (m^{\text{US}})^2 } +3 \kappa_3}{2 \lambda_\phi^{\rm US}  } \sim \frac{ \kappa_3}{ \lambda_\phi^{\rm US}  } \;, \quad \textrm{ with } \kappa_3 \sim \frac{(g^{\text{US}}_{f})^3}{16 \pi} \;,
\end{align}
where we have only included the dominant $(g^{\text{US}}_{f})^3$ contribution in the trilinear coupling. Note that we can neglect the square root in the above formula precisely because $\kappa_3 \gtrsim  \sqrt{\lambda_\phi^{\rm US}}  m^{\text{US}}$ is the condition for the second minimum to appear in the parameter space relevant for a strong first order transition. From the simple order-of-magnitude expression above, we see that in order to push the broken vacuum to larger field values -- and thus enhance the barrier between the true and the broken vacua -- one can either decrease $\lambda_{\text{US}}$ or increase the trilinear term. This can however be achieved up to the maximum splitting between both terms underlined above, since when $g_f^4$ matching corrections to $\lambda_{\rm US} $ begin to dominate, $ \langle \phi\rangle_{T_c} \propto 1/g_{f}^\text{US} $ and increasing the gauge coupling eventually weakens the phase transition.  We illustrate this effect based on our complete numerical framework in Fig.~\ref{fig:VEVingf}.

\begin{figure}[t]
\centering
\includegraphics[width=0.49\linewidth]{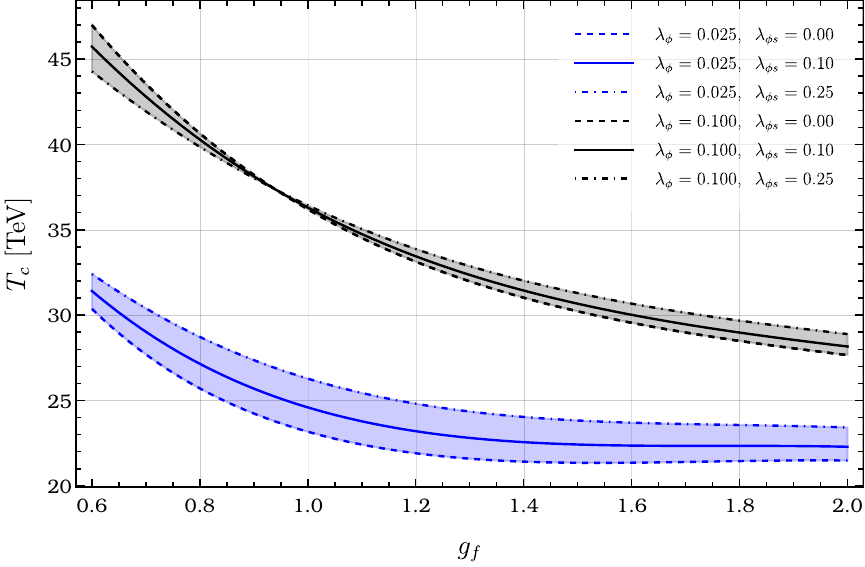}
\includegraphics[width=0.49\linewidth]{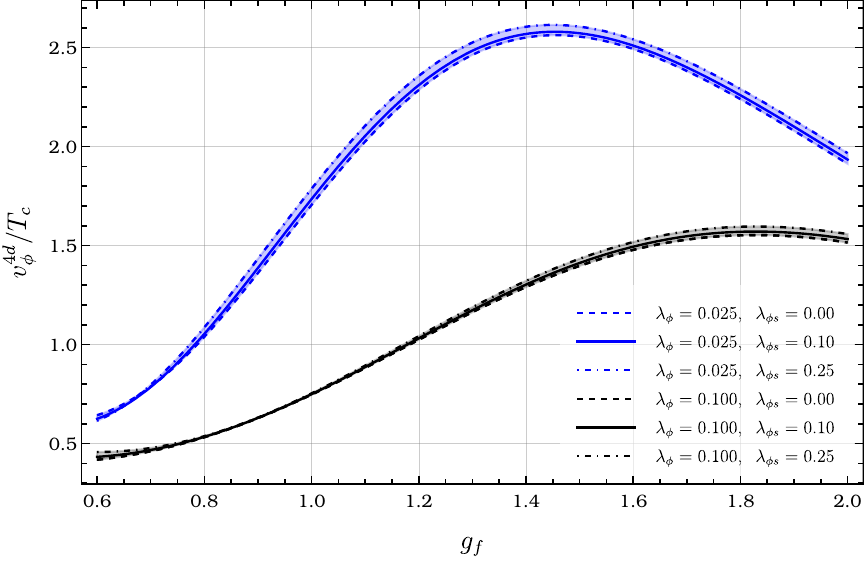}
\caption{(Left) Evolution of the critical temperature from the complete NNLO estimation of \dralgo. (Right) for the 4d VEV, we plot the corresponding ratio of VEV over critical temperature indicating the strength of the FOPT.
}
\label{fig:addVEVingf}
\end{figure}

Finally, while this is to some extent independent of the DR procedure, the one-loop RGE evolution in the 4d theory that we consider between the initial scale $v_0$ and the hard scale $\pi T$ where we match to the 3d theory also tends also to prevent large hierarchies between $\lambda$ and $g_f$. This can be straightforwardly seen from the one-loop beta-function:
\begin{align}
   \beta_\lambda =  \frac{3}{(4\pi) ^2}  \left(\frac{3}{8} g_f^4- 3  g_f^2 \lambda_\phi + 2 (\lambda_{\phi s}^2 +4 \lambda_\phi ^2 ) \right) \ .
\end{align}
\noindent We have further explicitly checked that in the parameter range considered in the rest of this work no Landau poles occurred.

\par For order-one gauge coupling $g_f$, we also have   $v_{\phi} / T_c \gtrsim 1$, and thus a FOPT \cite{Patel2011_GaugeIndependence} with a potentially sizeable GW signal from this model falls within the observational window. With this promising result in place, we proceed with the investigation of the thermodynamic properties of this model.

\subsection{Thermal parameters computed from the phase transition \label{subsec:PTobs}}
\par The thermal parameters that help characterise the phase transition are computed from the effective potential and the corresponding Euclidean action. In particular, we highlight
\begin{itemize}
    \item the \enquote{nucleation temperature}, $T_N$, considered an indicator of the onset of the phase transition and, in the absence of supercooling, as a proxy for the characteristic temperature upon which the hydrodynamic processes leading to GW production depend \cite{Athron2022_Subtleties};
    \item the \enquote{phase transition strength} evaluated at the nucleation temperature, $\alpha_N $: the latent heat of the phase transition, normalised with respect to the radiation energy of the plasma $\rho_R$ at the nucleation time in the symmetric phase (i.e. outside the bubbles) \cite{EllisLewicki:2019oqb,Giese:2020znk};
    \item a measure of the rate of change of the nucleation rate normalised with respect to the Hubble rate at the nucleation temperature, $\beta/H_N$, which conveys the \enquote{inverse duration} of the phase transition.
\end{itemize}

\paragraph*{Nucleation temperature and $\beta/H$.}
The precise definitions of each of the above thermal parameters vary within the community, and often depend on the objectives and conventions adopted in different studies. This is due in part to the fact that thermal parameters are evaluated at a single \enquote{transition temperature}, $T_*$; the correct milestone to which this temperature should correspond depends on the subtleties of the phase transition scenario under consideration.  A common simplifying choice, which we also follow here, is to express all quantities with respect to $T_N$. 
The nucleation temperature has been used extensively in the literature to characterise the onset of the transition and, provided phase transitions begin sufficiently quickly, should not be too distinct from the temperature at which bubbles collide \cite{Athron2023_GWreview}. However, in the presence of supercooling, 
one can use instead the \enquote{percolation temperature}, $T_P$, defined as the temperature at which a fraction $1-1/e \simeq 0.63$ of the universe has transitioned to the broken phase \cite{Croon:2020_ThermalResum,Brdar:2025gyo}. This definition ensures that the thermal parameters are directly related to the dynamics of the transition, irrespective of whether the specific bubble nucleation criteria are satisfied \cite{Caprini:2019_GW-PT-LISA,Enqvist:1991xw}.
 In practice, the two scales are numerically similar when the transition does not undergo strong supercooling with $\alpha_N \gtrsim 1$, so that $T_P \approx T_N$ remains a good approximation \cite{Athron2022_Subtleties}. Since the model studied here does not undergo supercooling, we opt to use $T_N$ as our reference temperature.\footnote{In the context of strong supercooled phase transitions, the approximate equivalence assumed in the literature between nucleation, percolation, and completion temperatures breaks down, as discussed in Ref. \cite{Athron2022_Subtleties}. Supercooling requires that the nucleation temperature be much lower than the symmetry-breaking scale, which in turn leads to a very strong FOPT characterised by $\alpha_* \gg 1$ \cite{Badger:2022nwo}. In fact, $\Delta V > \rho_{\rm rad}$ for supercooled phase transitions, which dilutes the pre-existing thermal bath due to a brief vacuum-dominated period. Bubble walls therefore encounter negligible friction and attain runaway behaviour (see Ref. \cite{Cataldi:2024pgt}).} 

\par
$T_N$ is usually described as the temperature at which, on average, one bubble per horizon volume is created. Using the nucleation time, $t_n$, and the time $t_c$ at which the critical temperature, $T_c$, is first reached, we can write the implicit expression for $T_N$ as
\begin{equation} \label{eq:Tn}
\int^{t_n}_{t_c} dt \frac{\Gamma (t) P_f (t) }{H(t)^3} \approx \int^{T_c}_{T_N} \frac{dT}{T} \frac{\Gamma (T )}{H(T)^4} =1 \;,
\end{equation}
\noindent where $P_f (t)$ is the \enquote{false vacuum fraction}, which conveys whether the phase transition completes (specifically, it is the probability that a random point lies in the false vacuum, i.e. is not enveloped by at least one bubble). Under the assumption that there is little to no supercooling ($T_c \approx T_N$) and that bubble nucleation begins swiftly ($\beta/H_N \gg 1$), we can approximate $P_f (t \leq t_n) \approx 1$:
in this scenario, many small bubbles nucleate before a significant fraction of the Universe is converted to the true vacuum, thereby ensuring that the phase transition does complete \cite{Athron2023_GWreview}.

\par Here, $\Gamma(T)$ is the bubble nucleation rate per unit volume, which for thermally activated tunnelling is given by \cite{Linde:1981zj},
\begin{equation}
    \Gamma(T) \simeq T^4 \left(\frac{S_3(T)}{2\pi T}\right)^{3/2} \exp\left(-\frac{S_3(T)}{T} \right),
\end{equation}
with $S_3$ serving as the 3d Euclidean action of the bounce solution. In the radiation-dominated early Universe, the Hubble parameter is given by
\begin{equation}
  H^2(T) = \frac{1}{3M_\text{Pl}^2} \left( \frac{\pi^2}{30} g_* T^4 + \Delta V(T) \right),
\end{equation}
where $M_\text{Pl} = 2.435 \times 10^{18}\,$GeV is the reduced Planck mass, $g_*$ is the effective number of relativistic degrees of freedom in the thermal bath in the symmetric phase, and $\Delta V(T) = V_{\rm eff}(0, T) - V_{\rm eff}(v_{\phi}, T)$ is the vacuum energy difference between the symmetric and broken phases.

\par For phase transitions at the electroweak scale (or for radiation-dominated epochs), the integral condition \eqref{eq:Tn} is well approximated by \cite{Caprini:2019_GW-PT-LISA,Guth:1979bh,Guth:1981uk}:
\begin{equation}
\frac{S_3(T_N)}{T_N} \approx 4 \log \left( \frac{T_N}{H(T_N)} \right)
\approx 140\;.
\end{equation}
While the value of $140$ should, strictly-speaking, be modified for a temperature-dependent criterion (see e.g. Ref.~\cite{Ellis:2018mja}), the logarithm implies that this criterium moves at most to $\sim 100$ (for a $6$ orders of magnitude shift in nucleation temperature). Given the very strong dependence of the tunnelling action on the temperature, we checked that this only modified the phase transition strength $\alpha_N$ by a $15 \%$ shift at most, far below our theory precision due to the NNLO nature of our effective potential calculations~\cite{EkstedtSchichoTenkanen:2024etx}. As such, we will neglect this temperature dependence in the remainder of this work.

\par Given $T_N$ and $S_3$, the inverse phase transition duration can be expressed as 
\begin{align} \label{eq:betaH}
\frac{\beta}{H_N}=T \left.\frac{d \left(S_3 (T)/T \right)}{dT}\right|_{T=T_N}\,,
\end{align}
where $H_N$ is the Hubble parameter at $T_N$.

\vspace{0.2cm}
We obtained throughout our numerical results the bubble action via the bounce method. We find the scalar field bounce solution along with the corresponding effective action using the Mathematica package, \texttt{FindBounce} \cite{FindBounce}, with our thermal effective potential as inputs, assuming that the LQ would not acquire a colour-breaking VEV based on the constraints found in Sec.~\ref{sec:su2}.\footnote{For the strongest phase transitions found in this work, with $\alpha_N \sim \mathcal{O} (0.1)$, \texttt{FindBounce} struggles to find numerically the exit point of the bounce solution; we thus implemented an adaptative strategy in the spirit of the approach used in Ref.~\cite{Brdar:2025gyo}.}

\paragraph*{Phase transition strength.}
\par In the previous section, we mentioned that the ratio of the scalar field VEV of the true vacuum over the critical temperature serves as an indicator of the strength of the phase transition. More quantitatively, we employ $\alpha_N \equiv \alpha (T=T_N)$, the latent heat of the phase transition (evaluated at nucleation temperature), normalised with respect to the radiation energy of the plasma $\rho_R$  in the symmetric phase \cite{EllisLewicki:2019oqb,Giese:2020znk}. 

Following the conventions of Ref. \cite{Kamionkowski:1993fg}, $\alpha$ can be described as the ratio between the \enquote{vacuum} energy and the radiation energy density in the symmetric phase. The vacuum energy $V_{\rm vac}$ is defined as in Ref.~\cite{Kamionkowski:1993fg} from the trace anomaly, 
\begin{equation}
\theta = \epsilon - 3p \;,
\end{equation}
as a quarter of the difference between the symmetric and broken phases, 
\begin{equation}
\Delta V_{\rm vac} = \frac{1}{4} (\theta_{\rm sym} - \theta_{\rm bro}) \;.
\end{equation}
Following this convention, the trace anomaly vanishes in the symmetric phase. In the broken phase, we then have 
\begin{equation}
\label{eq:VacEnergy}
 \theta_{\rm bro} =  -T \frac{d}{dT} V_{\rm bro}
 + 4 V_{\rm bro}
 \;.
\end{equation}
We can then obtain the expression for the strength of the phase transition, evaluated at $T=T_N$, the dimensionless parameter $\alpha_N$, 
\begin{equation} \label{eq:alphaN}
    \alpha_N \equiv \frac{\Delta V_{\rm vac}}{3aT^4} 
    \bigg \vert_{T=T_N} = \frac{1}{\rho_R} \left[ \Delta V_{\rm eff} (\phi,T) - \frac{T}{4} \Delta \frac{d V_{\rm eff} (\phi,T)}{d T} \right]_{T = T_N},
\end{equation}
where $3 a T^4 = \rho_R$ for $\rho_R = \pi^2 g_* T^4/30$. Here, $V_{\rm sym} \equiv V_{\rm eff}(0,T)$ and $V_{\rm bro} \equiv V_{\rm eff}(v_{\phi},T)$ denote the effective potential evaluated in the symmetric and broken phases, respectively \cite{EkstedtSchichoTenkanen:2024etx,Athron2023_GWreview}.\footnote{Note that the DR method, as implemented in \texttt{DRalgo}, only gives us the NLO matching conditions between the hard and the soft EFT in the vicinity of the symmetric vacuum \cite{DRalgo}; this implies that the NNLO contribution is only partially applicable in the broken phase~\cite{EkstedtSchichoTenkanen:2024etx}.}

\par The effective number of relativistic degrees of freedom in the true vacuum phase is given by $g^{\rm eff}_*$. Assuming that the relativistic particles are at thermal equilibrium,

\begin{align} \label{eq:DoF}
    g_* (T) &= \sum_B g_B + \frac{7}{8} \sum_F g_F \nonumber \\
   \Rightarrow \quad g_*^{\rm eff} & =   (28 + 1 + 12  + 9  ) + \frac{7}{8} ( 90 ) \nonumber \\
    &\approx 129 \;.
\end{align}
\noindent The SM contributes 28 bosonic and 90 fermionic degrees of freedom, such that $g_*^{\rm SM} (T \gtrsim 200 \;{\rm{GeV}})$ = 106.75). As summarised in Table \ref{tab:model}, the $\Phi$ field and the $S$ LQ contribute 1 and 12 degrees of freedom, respectively, with 9 degrees of freedom from the 3 massive vector flavour bosons and their 3 polarisations. The heavy fermions decouple from the thermal plasma and therefore do not contribute, leaving us to account for 22 new degrees of freedom.

\paragraph*{Numerical results and benchmark points}

\par 
Qualitatively, $\alpha_N \sim {\mathcal{O}}(0.01)$ denotes weak transitions, 
$\alpha_N \sim {\mathcal{O}}(0.1)$ corresponds to intermediate transitions, and $\alpha_N \sim {\mathcal{O}}(1)$ indicates a strong transition \cite{Athron2023_GWreview}. In order to illustrate the impact of the various aspects of our models on the thermal parameters, we have introduced five benchmark points (BPs), presented in Table \ref{tab:BPparams}. BP1 and BP2 focus on the simplest model with only a flavour $SU(2)_f$ gauge group and the $\phi$ doublet participating in the phase transitions. BP3, BP4, and BP5 instead include the $12$ degrees of freedom of the $S$ LQ with different masses and couplings to $\phi$.

\begin{figure}[t]
        \centering
        \includegraphics[width=0.65\linewidth]{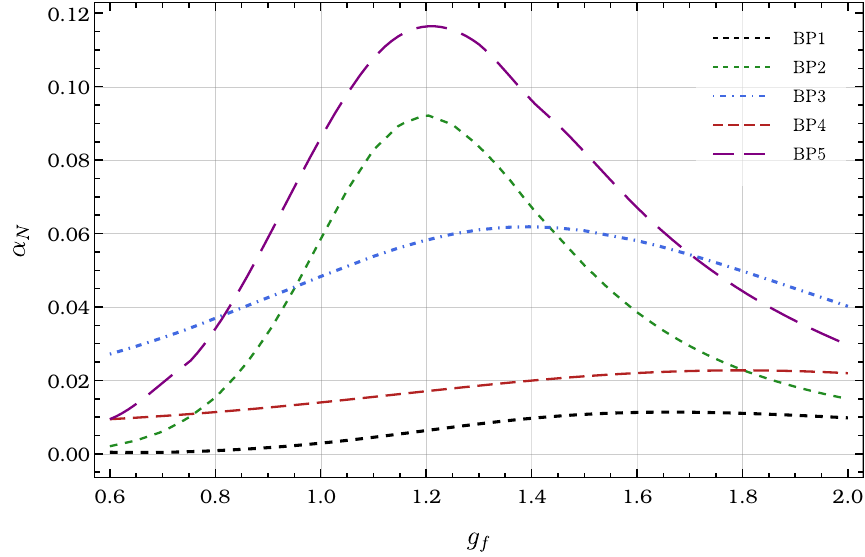}
        \caption{\textit{Evolution of $\alpha_N$ with $g_f$ for each of the benchmark points ($v_0=50$ TeV).
        }}
        \label{fig:alphafit}
    \end{figure}

\begin{figure}[t]
        \centering       
        \includegraphics[width=0.65\linewidth]{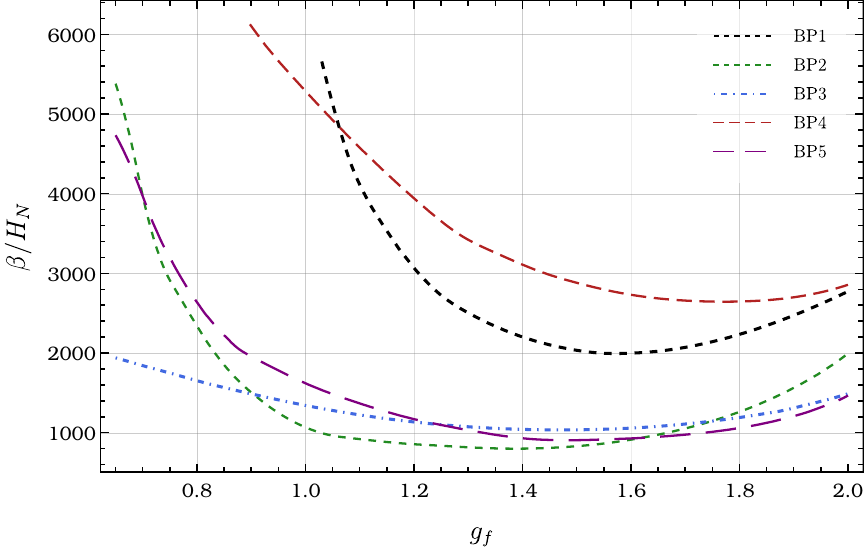}\caption{\textit{Evolution of $\beta/H_N$ with $g_f$ for each of the benchmark points ($v_0=50$ TeV). 
        }}
        \label{fig:betaHTN}
    \end{figure}

We present in Fig.~\ref{fig:alphafit} and Fig.~\ref{fig:betaHTN} the evolution of the thermal parameters $\alpha_N$ and $\beta/H_N$ for the benchmark points presented in Table~\ref{tab:BPparams} as function of the flavour gauge parameter $g_f$. We first observe that while all five lead to a first-order phase transition $(v_{\phi}/T_c > 1$), only BP2, BP3 and BP5 reach $\alpha_N$ values in the $\sim 0.05$ range. As explained previously, the phase transition strength typically reaches a maximum for $g_f\sim 1$, although the precise value depends on the tree-level value for the quartic coupling $\lambda_\phi$. 
 
We see from these figures and from Table~\ref{tab:BPparams} that $\alpha_N$ and $\beta/H_N$ are related: stronger phase transitions complete over a prolonged duration, i.e. large values of $\alpha_N$ correspond to small values of $\beta/H_N$, while weaker phase transitions proceed rather quickly with larger $\beta/H_N$.

\begin{table}[t]
    \centering
    \setlength\tabcolsep{0.1cm}
    \def\arraystretch{1.5}
    \begin{tabular}{@{}C|CCCC|CCCC@{}}
    \hline\noalign{\smallskip}
& m_S/v_0 & \lambda_{\phi} &  \lambda_{\phi s} &  g_f & T_c  \text{ [TeV]} & T_N \text{ [TeV]} \; (\ref{eq:Tn}) &  \alpha_N  \; (\ref{eq:alphaN}) & \beta/H_N \; (\ref{eq:betaH}) \\
\hline 
 \text{BP1} & 1.00 & 0.025 & 0.00 & 1.5 & 21.28 & 19.06 & 0.011 & 1886 \\ 
\text{BP2} & 1.00 & 0.005 & 0.01 & 1.0 & 16.95 & 12.34 & 0.055 & 1123  \\ 
\text{BP3}& 0.25 & 0.025 & 1.00 & 1.0 & 23.61 & 19.50 & 0.048 & 1339 \\ \
\text{BP4} & 0.10 & 0.100 & 1.50 & 1.5 & 25.76 & 24.09 & 0.021 & 2850\\
\text{BP5} & 0.50 & 0.005 & 0.25 & 1.5 & 17.37 & 13.76 & 0.085 & 884 \\
\hline\noalign{\smallskip}
    \end{tabular}
        \caption{\textit{Thermal parameters computed for each of the five benchmark points, where the $S$ LQ is kept at the US scale. We set $m_{\phi}^2 = \lambda_{\phi}v_0^2\,, \; m_s^2 =v_0^2, \; \lambda_{s}=0.5,$ and $v_0=50$. See Appendix \ref{app:BPparams} for additional parameters.}}
    \label{tab:BPparams}
\end{table}

\par From these parameters, we can also gain some insight into the expected GW signal. The GW signal is predicted to be enhanced for larger $\alpha_N$; a higher $T_N$ is associated with a peak in the lower frequency range; since delaying the phase transition leads to a larger $v_{\phi}/T_c$, smaller values of $\beta/H_N$ improve the detection prospects \cite{Caprini:2019_GW-PT-LISA}.

\begin{figure}[t]
\centering
\includegraphics[width=0.7\linewidth]{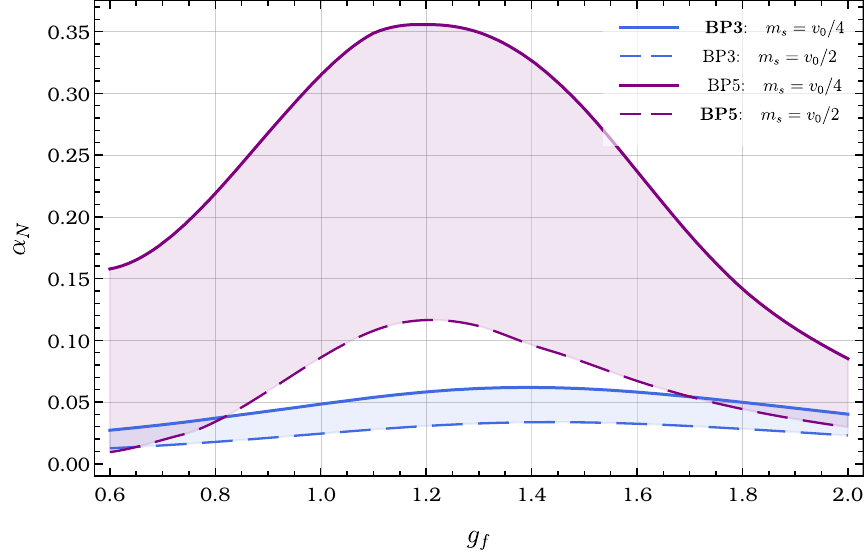}
\caption{\textit{The effect of $m_s$ on the phase transition strength for BP3 and BP5. Boldface in the legend corresponds to the choice of $m_s$ in Table \ref{tab:BPparams}. }}
\label{fig:alphalm0-025} 

\end{figure}

It is clear that small $\lambda_\phi$ (at or below the percent level) leads to intermediate FOPT in our scenarios, with larger quartic typically yielding only weak ones. However, in the case of BP3, we see that a $\lambda_\phi = 0.025$ can still lead to intermediate FOPT due to the interaction between the relatively light LQ and $\phi$ for this BP. 
Indeed, in the presence of the LQ, we can use the scalar mixing parameter $\lambda_{\phi s}$ effectively to reduce the US quartic couplings as shown in Eq.~\eqref{eq:UScouplings} or equivalently have it participate directly in the trilinear barrier in the effective potential when they are light enough. We illustrate this effect in Fig.~\ref{fig:alphalm0-025}, where we have further shown the effect of varying by $2$ the mass of the LQ with respect to the reference scale $v_0$. We obtain a correspondingly linear increase in the phase transition strength, confirming the importance of having additional particles around the flavour-breaking scale to obtain a FOPT strong enough to generate eventually a significant GW spectrum. Furthermore, we also tested that using a larger representation for the $S$ LQ -- as a simple proxy to increase the number of degrees of freedom in this field -- further increased the phase transition strength. With around $36$ new degrees of freedom (corresponding, for instance, to having $S$ in a sextet of colour and a triplet of $SU(2)_f$), we could obtain for BP3 an intermediate $\alpha_N \sim 0.1$, while roughly doubling the possible $\lambda_\phi$ values at levels closer to the SM Higgs reference value at these multi-TeV scales.

To conclude this section, it is important to stress that while the Figures above have been realised for a benchmark value of $v_0 = 50$ TeV, our results can be straightforwardly adapted to any other desired value by the rescaling 
\begin{align}
    T_N \to \left( \frac{v_0}{50 \, \textrm{TeV}}\right) \, T_N \  \qquad \, \ T_c \to \left( \frac{v_0}{50 \, \textrm{TeV}}\right) \, T_c \ ,
\end{align}
while leaving all the other relevant thermal parameters ($\beta/H$ and $\alpha_N$) unchanged. This simple scaling originates from: (1) the fact that the SM effective number of degrees of freedom does not evolve further once above the TeV scale, and (2) we have neglected the variation in temperature in our nucleation temperature criterium (i.e. in each case, the calculated temperature $T_N$ is approximately constant throughout the duration of the phase transition). All in all, $v_0$ is therefore the only absolute energy scale left in our problem, with all the other dimensionful parameters defined after it, thus leading to the above linear scaling. As we have discussed previously, we expect both assumptions to hold consistently given the precision of our effective potential calculation for the energy ranges accessible to future GW observatories (up to $10^7$ GeV for the Einstein Telescope).

\section{GW signatures of $\sux$ models}
\label{sec:gw}
 
\par We anticipate a strong FOPT when the sphaleron decoupling condition and the phase transition strength, respectively, are sufficiently large, \textit{viz.} $v_{\phi}/T_c, \;\alpha_N \sim \mathcal{O}(1)$. However, to predict the amplitude and frequency of the GW signal generated by this FOPT, we must consider the behaviour of the nucleated bubbles of the broken phase, the behaviour of the surrounding plasma, and the interactions between the bubble walls and the plasma.  

\par These discussions of the dynamics of the phase transition and the resulting GW signal rely heavily on considerations of the energy budget of the universe \cite{Espinosa:2010_EnergyBudget}. This is because upon its nucleation, each bubble of the broken phase that is large enough to begin growing will expand at an accelerated pace into the plasma of hot particles. The surrounding plasma resists this bubble wall expansion, exerting a pressure against the moving wall. After some time, an equilibrium may be established between the two opposing forces, and as a result, the bubble wall continues to expand in a steady state with a constant terminal velocity. 

\par Bubble wall speed affects the magnitude of the GW signal, where faster-moving bubble walls generally produce larger GW amplitudes. The last fifteen years have seen substantial progress in the modelling of the wall-plasma interactions and relativistic hydrodynamics, involving analytical study and numerical simulation from a number of highly active groups (see reviews \cite{Weir2017_GWsFOPTreview,Caprini:2019_GW-PT-LISA,Athron2023_GWreview,vandeVis:2025efm}). In this section, we outline how we model the bubble dynamics according to the model-agnostic framework of Refs \cite{Espinosa:2010_EnergyBudget, EllisLewicki:2022lft}, while also incorporating the influence of wall-plasma friction \cite{Lewicki:2021pgr} and the infrared cut-off for bosonic degrees of freedom \cite{DeCurtis:2024hvh}.

\par To simplify the hydrodynamic description of the plasma surrounding the bubble, a relativistic fluid approximation can be used. In this context, the plasma's thermodynamic quantities, such as energy density $e$ and pressure $p$, are governed by what is known as the \enquote{bag equation of state} \cite{Espinosa:2010_EnergyBudget,Megevand:2013hwa} that we will present below. 

\par Another assumption is that of the scalar field profile. When the bubble is sufficiently large for the steady-state solutions, we can interpret the problem as one-dimensional, justifying a planar approximation. For this reason, we model the scalar $\phi$ field by the $\tanh$ function \cite{Friedlander:2020tnq},
\begin{equation} \label{eq:tanh}
\phi (z) = \frac{1}{2} \phi_N \left[\tanh \left( \frac{z}{L_w} \right) +1 \right] \;.
\end{equation}
\noindent Such an expression corresponds to the instanton solutions for bubble nucleation, allowing scalar fields to interpolate continuously across the bubble \cite{Lewicki:2021pgr}. The quantities $\phi_N$ and $L_w$ denote the minimum of the scalar field evaluated at $T_N$ and the wall thickness, respectively. We work in the rest frame of the bubble wall and consider the direction of the wall propagation to be $+z$.

\subsection{Wall-plasma interactions}

\par The effective thermal potential $ V_{\rm eff}$ calculated in the previous section is the free-energy density of the plasma. Outside of the wall, where the plasma is in equilibrium, we may thus use simple thermodynamics to describe its state. However, when crossing the wall, particles may leave momentarily their thermal equilibrium distributions. We will describe the latter by an effective \enquote{friction} term following Refs.~\cite{Espinosa:2010_EnergyBudget, EllisLewicki:2022lft}.

\paragraph{Plasma in the broken and symmetric phase} Assuming first that the plasma is always locally at equilibrium, we may then characterise the metastable false vacuum by $ \mathcal{F}_{+}(T)= V_{\rm eff}(\phi_+,T)$ and the stable true vacuum by $\mathcal{F}_{-}(T)= V_{\rm eff}(\phi _{-},T)$.\footnote{We maintain the convention that a plus sign refers to the symmetric phase and a minus sign to the broken phase.} Correspondingly, we define the pressure in each phase as
$p_\pm=-\mathcal{F}_\pm$, the entropy density as $s_\pm=dp_\pm/dT$, and the energy density as $e_\pm=Ts_\pm-p_\pm$. We can finally define the latent heat from the entropy densities' discontinuity at $T=T_c$, and given by $L=T_c[\mathcal{F}_-'(T_c)-\mathcal{F}_+'(T_c)]$.

\par The hydrodynamics of the plasma surrounding the bubble are well-approximated as a perfect relativistic fluid. In this context, plasma thermodynamic quantities, such as energy density $e$ and pressure $p$, are governed by the \enquote{bag equation of state} \cite{Espinosa:2010_EnergyBudget,Megevand:2013hwa}, with the speed of sound $c^2_s = (dp/dT)/(de/dT) = 1/\sqrt{3}$.\footnote{We assume that the propagation velocity of sound in the plasma is approximated here as equal on both sides of the wall i.e. $c_s = c^+_s = c^-_s$.} In the unbroken phase,
\begin{equation}
\label{eq:pplus}
    p_+ = \frac{1}{3}a_+T_+^4 - \varepsilon \;, \qquad e_+ = a_+T_+^4 + \varepsilon \;,
\end{equation}
\noindent where $\varepsilon$ is the vacuum energy difference between the two minima. In the broken phase,
\begin{equation}
\label{eq:pminus}
    p_- = \frac{1}{3}a_-T_-^4 \;, \qquad e_- = a_-T_-^4 \;.
\end{equation}
\noindent Here, we have introduced
\begin{equation} 
    a_{\pm} = \frac{\pi^2}{30} \sum_{{\rm{light}} \; i} g_i(T)  \;,
\end{equation} 
\noindent where $g_i(T)$ refers to the effective number of relativistic degrees of freedom (c.f. Eq. \eqref{eq:DoF}, corresponding to degrees of freedom after symmetry breaking). 
Since in the following we will vary the LQ masses around the phase transition scale, they cannot be counted as purely relativistic degrees of freedom in the symmetric nor in the broken phase, and we do not include them in the above ratio; we include instead only the new bosons. As such, we count $a_+ \sim 117 \pi^2/30$ in the symmetric phase. For the broken phase, we use the SM value, $a_- \sim 107 \pi^2/30$ \cite{Espinosa:2010_EnergyBudget}. The main effect of this choice is to move marginally the boundary between the first- and the second-order phase transition (note that in the case where more degrees of freedom were to be included in the full theory, this point ought to be treated more precisely).\footnote{In the SM $g_*^{\rm SM} (T \geq 200 \rm{GeV}) = 106.75$. The top quarks are the first to decouple, such that 12 degrees of freedom are lost to the fermions. As such, $g_{SM} (T \sim 30 \rm{GeV}) =96.25$. } 

\paragraph{In-plasma scalar equations of motion}  The bubble expansion is driven by the difference in free-energy density across the bubble wall (\textit{viz.} $\mathcal{F}_{-}(T)$ within the bubble, in the broken phase, and $ \mathcal{F}_{+}(T)$ outside the bubble, in the symmetric phase). As such, to capture the bubble dynamics, we must consider the energy-momentum tensor of the field driving the bubble expansion,
\begin{equation}
\label{eq:TmunuPhi}
T_{\mu\nu}^{\phi} = \partial_{\mu}\phi \partial_{\nu}\phi -g_{\mu\nu} \left[ \frac{1}{2} \partial_{\rho}\phi \partial^{\rho}\phi - V_{\rm eff} (\phi,T)\right] \;,
\end{equation}
as well as that of the surrounding plasma,
\begin{equation}
\label{eq:TmunuPlasma}
 T_{\mu\nu}^{ \rm pl} = \sum_{i} \int \frac{d^3 p}{(2\pi)^3 E_{i}} p_\mu p_\nu f_{i}(p_{\mu},x) \;, 
\end{equation}
\noindent where the sum is carried out over the species in the plasma and $f_{i}(p_{\mu},x) $ is the distribution function for each species. To encode the behaviour of this wall-plasma system, we shall consider the distribution function as
\begin{equation} \label{eq:distfunc}
    f_{i}(p_{\mu},x) = f^{\rm eq}_{i} (p_{\mu},x) + \delta f_{i} (p_{\mu},x) \;,
\end{equation}
\noindent where the plasma is locally in equilibrium when $f_{i}(p_{\mu},x) = f^{\rm eq}_{i} (p_{\mu},x)$ and deviations from thermal equilibrium are denoted by $\delta f_{i} (p_{\mu},x)$. Here, 
\begin{equation}
    f^{\rm eq}_{i} (p_{\mu},x) = \frac{1}{ \exp \left[(p_{\mu} u^{\mu} )/T \right] \pm 1} \;,
\end{equation}
\noindent with $+$($-$) for fermions (bosons). With Eq. \eqref{eq:distfunc} in mind, we can rewrite Eq. \eqref{eq:TmunuPlasma} as a sum of its equilibrium and out-of-equilibrium contributions,
\begin{align}
T_{\mu\nu}^{\rm pl} &= T_{\mu\nu}^{\rm eq} + T_{\mu\nu}^{\rm out} \;, \nonumber \\
 T_{\mu\nu}^{\rm eq} &= \sum_{i} \int \frac{d^3 p}{(2\pi)^3 E_{i}} p_\mu p_\nu f^{\rm eq}_{i}(p_{\mu},x) \;, \nonumber \\
 T_{\mu\nu}^{\rm out} &= \sum_{i} \int \frac{d^3 p}{(2\pi)^3 E_{i}} p_\mu p_\nu \delta f_{i}(p_{\mu},x)
\;. 
\end{align}

\noindent By conservation of energy-momentum, 
\begin{equation} \label{eq:TmunuconserveOOE}
\partial^\mu T_{\mu\nu} = \partial^\mu T_{\mu\nu}^{\phi} + \partial^\mu T^{\rm eq}_{\mu\nu} + \partial^\mu T_{\mu\nu}^{\rm out} = 0 \;,
\end{equation}
which can equivalently be written as
\begin{equation} \label{eq:EoMphiLewicki}
E \equiv \Box \phi + \frac{\partial V_{\rm eff}(\phi,T)}{\partial \phi} - \mathcal{K}(\phi) = 0 \;,
\end{equation}
where we included all out-of-equilibrium contributions into $\mathcal{K}(\phi)$, with
\begin{align}
 \mathcal{K}(\phi) = -   \sum_{i} \frac{d m^2_i}{d \phi} \int \frac{d^3 p}{(2\pi)^3 } \frac{\delta f_{i}(p_{\mu},x)}{2 E_{i}} \;.
\end{align}

\noindent When the bubble is large enough to justify a planar approximation (c.f. Eq. \eqref{eq:tanh}), the conservation equation simply reads in the wall frame as 
\begin{equation} \label{eq:TmunuconserveWallz}
    \partial_z T_{zz} = \partial_{z} T_{z0} = 0 \;,
\end{equation}
\noindent where the bubble wall and the fluid velocities are aligned in the $z$ direction. 
\par Based on the above, if we impose the planar approximation, the equation of motion for the scalar field $\phi$ including both in-equilibrium and out-of-equilibrium contributions \cite{MooreProkopec:1995si,Espinosa:2010_EnergyBudget} can be written as  \begin{equation} \label{eq:EoMphiEspinosa}
\partial^2_z \phi  - \frac{\partial V_{\rm eff}}{\partial \phi} +  \mathcal{K}(\phi)  = 0 \;,
\end{equation}

\par Since the hyperbolic tangent from Eq. \eqref{eq:tanh} is only an approximation to the proper result of Eq. \eqref{eq:EoMphiLewicki} that cannot be satisfied everywhere, we follow Ref.~\cite{Lewicki:2021pgr} in imposing instead that the first moments of this equation vanish, \textit{viz.}: 
\begin{equation} \label{eq:M1}
M_1 \equiv \int dz \; E \; \frac{\partial \phi}{\partial z} dz = 0 \;,
\end{equation}
\noindent corresponding to the total pressure acting on the bubble wall (which must vanish for steady-state solutions \cite{MooreProkopec:1995si}), and
\begin{equation}  \label{eq:M2}
M_2 \equiv \int dz \; E \; \frac{\partial \phi}{\partial z} \; \tanh \left( \frac{z}{L_w} \right) dz = 0 \;,
\end{equation}
\noindent which serves as a first constraint on the bubble wall shape and that will, in practice, fix its thickness~\cite{Lewicki:2021pgr}, $L_w$.
More precisely, to derive $L_w$, we neglect the out-of-equilibrium terms in the second moment and consider a constant temperature $T_N$.
Using $E \partial_z \phi dz = \partial_z T_{zz}^{\phi}$ for the constant temperature $T_N$, we can then rewrite Eq. \eqref{eq:M2} as 
\begin{align}
M_2 &= \int dz \; \frac{\partial T_{\phi}^{zz}}{\partial z} \;  \phi(z)   \nonumber \\
&= - V_{\rm eff}(\phi_N,T_N)\phi_N - \int dz \frac{1}{2} \left( \frac{\partial \phi}{\partial z} \right)^3 + \int dz V_{\rm eff}(\phi,T) \; \frac{\partial \phi}{\partial z} \;,
\end{align}
\noindent where we use integration-by-parts to obtain the second line. Since $\int dz (\partial_z\phi)^3/2 = \phi_N^3/30L_w^2$ when using the $\tanh$ approximation of Eq. \eqref{eq:tanh}, the bubble wall thickness can be computed as
\begin{equation} \label{eq:thickness}
L_w^2 = \frac{\phi_N^3}{30\left[ \int_{- \infty}^{+ \infty} V_{\rm eff}(\phi,T) d\phi   - V_{\rm eff}(\phi_N,T_N) \phi_N \right] } \;.
\end{equation}

\paragraph{Friction and out-of-equilibrium contributions}

The first moment equation, Eq.~\eqref{eq:M1}, encodes the requirement that the pressures balances on both side of the wall. While the equilibrium part can be derived straightforwardly, the out-of-equilibrium contribution is more challenging to obtain. We will thus rely on the effective framework of Ref.~\cite{Espinosa:2010_EnergyBudget}, where the out-of-equilibrium contribution in the wall frame is conveyed by the third term of Eq. \eqref{eq:EoMphiEspinosa}, and can be expressed as
\begin{equation} \label{eq:Pouteta}
P_{\rm out}^{\eta} = \int dz \, \partial_z\phi \;\mathcal{K} (\phi) ~\equiv~  (a_+ T_N^4)  \; \eta \; \langle v \rangle   \;.
\end{equation}
\noindent Here, $\eta$ is an effective friction parameter that is typically freely varied in hydrodynamics simulation, and the fluid velocity average across the wall (in the wall frame) is given by
\begin{equation}
\langle v \rangle = \frac{\int dz \; v \left( \partial_z \phi  \right)^2}{\int dz\left( \partial_z \phi  \right)^2} \approx \frac{1}{2} (v_+ + v_-)
\end{equation}
\noindent for the fluid velocity in front of, $v_+$, and behind, $v_-$, the wall. The above relation is the direct consequence of assuming that 
\begin{align}
     \mathcal{K}(\phi) \propto v \, T_N \partial_z \phi \ ,
\end{align}
with $v$ as the fluid velocity.

 \par  As demonstrated in Ref. \cite{Megevand:2009ut}, $\eta$ can be computed explicitly from the microscopic theory~\cite{MooreProkopec:1995si}, with the top quark providing the dominant contribution to the out-of-equilibrium behaviour. 

Instead, in our case, we anticipate that friction is dominated by the flavour gauge bosons. Due to the delicate infrared behaviour of the gauge bosons within the thermal bath already discussed in Section \ref{subsec:DRexplained}, these contributions must be handled with care.\footnote{Note that we did not include the LQ in the estimation of out-of-equilibrium effect at this point $-$ in part, since they are already partially massive before the phase transition, and in part because this delves into the precise particle physics aspects of the models. We leave this to future works.}

\par To do so, let us first return to Eq. \eqref{eq:EoMphiLewicki}, where the third term can be expressed as the friction measured in the wall reference frame as
\begin{equation} \label{eq:FrLewicki}
\mathcal{K} = \sum_{i} \frac{N_i}{2} \frac{d m^2_i}{d z} \int \frac{d^3 p}{(2\pi)^3 } \frac{\delta f_{i}(p,z)}{ E_{i}}  \;,
\end{equation}
\noindent for the degrees of freedom $N_i$, the mass $m_i$, and out-of-equilibrium contributions $\delta f_i$ of each species $i$. 
The behaviour of these non-equilibrium distributions of the massive particles in the plasma are modelled using the Boltzmann transport equations. In Ref. \cite{DeCurtis:2024hvh}, $\delta f_i$ is treated as perturbations around equilibrium, where their approach employs a multipolar expansion method and the application of functional spectral methods to solve the linearised Boltzmann equation. When considering non-linear contributions, it becomes apparent that the contribution of gauge bosons to the friction is an order of magnitude larger than naive scaling suggests \cite{Moore:2000wx}. This is because the Boltzmann equation fails to capture the dominating Landau damping and the screening effects that characterise the soft modes ($p \leq g T$). To address this behaviour of the gauge bosons, they solve the Langevin equation. We can express their result here for the out-of-equilibrium contribution from the flavour gauge bosons, 
\begin{equation}
\delta f_f = \frac{\pi m^2_{\rm D} \gamma_w v_w}{16 p E^3_f T_N} f^{\rm eq}_f(1+f^{\rm eq}_f) \frac{d \phi}{dz} \frac{d m^2_f}{d \phi}  \;,
\end{equation}
\noindent where the index $i=f$ indicates that the species in question is the flavour gauge boson. The integrated friction can be referred to as the pressure from the out-of-equilibrium contributions,
\begin{equation} \label{eq:Pout}
P^{m_f}_{\rm out} = \gamma_w v_w \frac{9 m^2_{\rm D} T_N}{32 \pi L_w } \int^1_0 \frac{1-x}{x} dx \;.
\end{equation}
\noindent Here, $x \equiv \phi(z)/\phi_N$ (c.f. Eq. \eqref{eq:tanh}), such that $x=0$ and $x=1$ correspond to $z=-\infty$ and $z=+\infty$, respectively. The evident infrared-divergence originates from the ultrasoft particles with non-zero mass in the symmetric phase, whose dynamics cannot be captured appropriately even by the Langevin equation. This is due to the treatment of gauge bosons as classical fields, which breaks down when the particle wavelength approaches that of the wall thickness, i.e. $\lambda_f \sim L_w.$ The resolution used in Refs~\cite{Moore:2000wx,DeCurtis:2024hvh} is the introduction of an infrared cut-off that effectively removes the contribution of ultrasoft particles in the symmetric phase. To accommodate the condition,
\begin{equation}
\lambda_f \ll L_w \quad \Rightarrow \quad m_f (v_{\phi}) L_w \gg 1 \;,
\end{equation}
\noindent we express the infrared cutoff as
\begin{equation}
x_{\rm IR} = \frac{1}{m_f (v_{\phi}) L_w} \;,
\end{equation}
for the mass of the flavour gauge boson in the broken phase, $m_f (v_{\phi}) $. Upon introducing this infrared cutoff to Eq. \eqref{eq:Pout}, and incorporating this in the comparison between Eqs. \eqref{eq:Pout} and \eqref{eq:Pouteta}, we can extract the expression for the model-dependent $\eta$,

\begin{equation} \label{eq:etahat}
    \eta =   r_v  \left( \frac{1}{L_w T_N} \right) \times \frac{1}{a_+}  \frac{ 9 m_{\rm D}^2}{32 \pi T_N^2} \left[ \log \left( \frac{1}{x_{\rm IR}} \right) - (1- x_{\rm IR} ) \right] \;,
\end{equation}
where we have defined the velocity ratio $ r_v ~\equiv~ \left( \gamma_w v_w/\langle v \rangle \right)$. In the following, we will assume this factor to be $r_v \sim 1$, as this will allow to use the matching equations across the wall directly as described in the next section. As we will see, in most hydrodynamical regimes we have $ v_w \sim \langle v \rangle $ and $\gamma_w \sim 1$ up to order one corrections, so that this assumption will hold except for the case of a runaway wall. In the later case, we follow ~\cite{Espinosa:2010_EnergyBudget} in that we do not include the Lorentz boost factor as it leads to an infinite increase of the friction, in contradiction with a proper out-of-equilibrium pressure calculation.

Altogether, we can now solve the first moment equation corresponding to the balancing of pressures on both side of the wall following Ref.~\cite{Espinosa:2010_EnergyBudget} as:
    \begin{equation}
      \alpha_+-\frac{1}{3}\left(1-\frac{a_-}{a_+}\right) = \eta \frac{\alpha_+}{\alpha_-} \langle v \rangle\, ,
        \label{eq:Pbalancing}
    \end{equation}
which will be used to find the wall position.

\subsection{Plasma hydrodynamics around the wall \label{subsec:hydro}}

\par To describe the plasma hydrodynamics, we assume that the system around the wall has reached local thermal equilibrium; this justifies the application of the framework introduced in Ref. \cite{Espinosa:2010_EnergyBudget}. In so doing, we are able to capture the hydrodynamics of the bubble wall velocity and the plasma fluid velocity in the different bubble expansion regimes: deflagration, detonation, and hybrid regimes in the case of steady-state solutions, and runaway for accelerating wall solutions.

\paragraph{Matching relations around the wall} To describe the plasma in the vicinity of the bubble wall, we come back to its stress-energy tensor:

\begin{equation}  \label{eq:TmunuEoM}
    T_{\mu\nu}^{\rm eq} =  w  u_\mu u_\nu  - p g_{\mu\nu}  \;, 
\end{equation}
where once again $w = e + p$ denotes the enthalpy, $p$ is the pressure of the fluid, and $e$ its energy density. Note that the constant $\phi$ background also contributes to this total pressure. The quantity $u_{\mu}$ denotes the four-velocity field of the plasma which is no longer at rest due to its interaction with the bubble wall. It can be expressed in terms of the 3d plasma velocity $\mathbf{v}$ as
\begin{equation}
    u = \frac{(1,\mathbf{v})}{\sqrt{1-\mathbf{v}^2}} = (\gamma,\gamma\textbf{v}) \;.
\end{equation}
\noindent We assume that the bubble wall velocity is constant, i.e. that there is no time dependence so that we are looking for steady-state -- stationary -- solutions. Now that the system has reached thermal equilibrium, the conservation of energy-momentum (c.f. Eq. \eqref{eq:TmunuconserveOOE}, before thermal equilibrium is reached) reads as 
\begin{equation} \label{eq:TmunuconserveEq}
\partial^\mu T_{\mu\nu} = \partial^\mu T_{\mu\nu}^{\phi} + \partial^\mu T_{\mu\nu}^{\rm eq} = 0 \;.
\end{equation}
\noindent Upon integrating these equations within the planar approximation (i.e. Eq. \eqref{eq:TmunuconserveWallz}) across the wall, we obtain
\begin{equation}
\label{eq:matching}
    w_+v_+^2\gamma_+^2 + p_+ = w_-v_-^2\gamma_-^2 + p_- , \qquad w_+v_+\gamma_+^2 = w_-v_-\gamma_-^2 \ .
\end{equation}
\noindent Following the earlier conventions, we denote the symmetric phase by \enquote{$+$} and the broken phase by \enquote{$-$}. Recall that the fluid ahead of the bubble wall corresponds to the symmetric (unbroken) phase, whereas the fluid behind (enclosed by) the bubble wall corresponds to the broken phase (see Fig. \ref{fig:sketch}).

\par We can also define
\begin{equation}
    \alpha_{+} \equiv \frac{\varepsilon}{a_{+} T_{+}^4}, 
    \qquad 
    r \equiv \frac{a_{+} T_{+}^4}{a_{-} T_{-}^4} = \frac{w_+}{w_-} \;.
\end{equation}
\noindent Here, $\alpha_{+}$ is the phase transition strength evaluated at $T_+$ (c.f. Eq. \eqref{eq:alphaN}) representing the ratio between the vacuum energy and the radiation energy density. This allows us to express the plasma velocity in the symmetric phase as
\begin{equation}
    v_{+} = \frac{1}{1+\alpha_{+}} \left[
    \left( \frac{v_{-}}{2} + \frac{1}{6 v_{-}} \right)
    \pm \sqrt{
    \left( \frac{v_{-}}{2} + \frac{1}{6 v_{-}} \right)^2
    + \alpha_{+}^2 + \frac{2}{3}\alpha_{+} - \frac{1}{3}
    }
    \right]
    \label{vplus}\;.
\end{equation}

\par From Eq. (\ref{vplus}), we see that for each value of $\alpha_+$, there are two solutions: the solution with the plus sign (i.e. upper branch) corresponds to a detonation, while the solution with the minus sign (i.e. lower branch) corresponds to a deflagration. Detonation presents as the most trivial case, in which $T_+ = T_N$ and $\alpha_+ = \alpha_N$. We note that for $\alpha_{+} > 1/3$, there is no deflagration solution. For deflagrations and hybrids, the plasma in front of the bubble wall is heated and accelerated. Stronger phase transitions result in higher velocities, and therefore thinner surrounding fluid shells. When the fluid shell disappears altogether, we switch to the detonation regime. The velocity at which this transition occurs is referred to as the \enquote{Jouguet velocity}, $v_J$. Note that Jouguet velocity can be defined directly from Eq. \eqref{vplus},
\begin{equation} \label{eq:vJ}
   \lim_{v_- \to 1/\sqrt{3}} v_+  = v_J \equiv \frac{1}{\sqrt{3}} \frac{1 + \sqrt{3 \alpha^2 + 2 \alpha}}{1 + \alpha} \;. 
\end{equation}

\paragraph{Regimes of bubble propagations}

The solutions to the hydrodynamic fluid equations can thus be classified into the aforementioned steady-state and accelerating family of solutions \cite{Espinosa:2010_EnergyBudget,Lewicki:2021pgr} according to the relationships between these plasma velocities and with the propagation velocity of sound in the plasma.

These regimes are characterised as follows:
 \begin{figure}[t]
     \centering
     \includegraphics[height=0.4\linewidth]{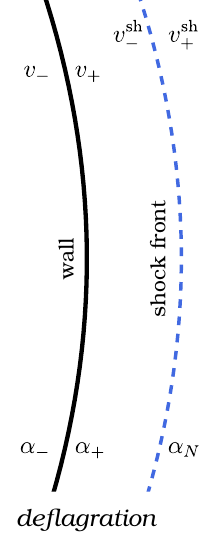}
     \hspace{1.5cm}
     \includegraphics[height=0.4\linewidth]{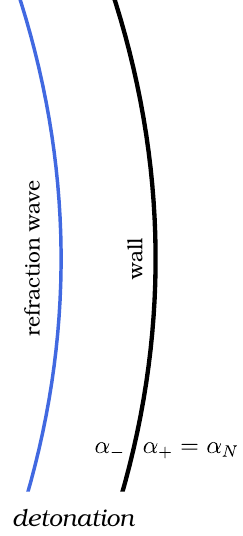}
     \hspace{0.8cm}
     \includegraphics[height=0.4\linewidth]{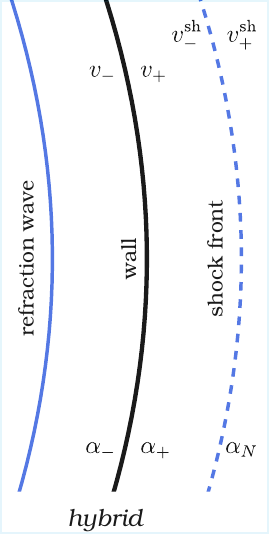}
     \caption{\textit{Sketch of the hydrodynamic regimes corresponding to the three steady-state solutions. }}
     \label{fig:sketch}
 \end{figure}
 \begin{itemize}
     \item  \textbf{Deflagration.} The plasma velocity behind the bubble wall vanishes ($v_- = v_w$), and is always smaller than the sound speed, i.e. $(v_- = v_w) < c_s$. As such, a shock wave forms in front of the subsonic wall. Thermodynamic quantities suffer discontinuities, which allows the solution $v_+$ to go to zero.
    \item  \textbf{Detonation.} In contrast, the plasma velocity in front of the bubble wall vanishes ($v_+ = v_w$), and is always larger than the sound speed, i.e.  $(v_+ = v_w) > c_s$. The bubble wall propagates supersonically into the unperturbed plasma, and the thermodynamic variables in front of the wall can be computed trivially.
\item \textbf{Hybrid.} A superposition of a deflagration and a detonation, in which the fluid behind the wall becomes sonic ($v_- = c_s$). The wall speed does not match the velocity behind or in front of the wall, but instead falls between the sound speed and the Jouguet velocity, $v_J$, defined in Eq. \eqref{eq:vJ}, i.e. $c_s < v_w < v_J$.
 \item \textbf{Runaway.} Plasma back-reaction is negligible and the wall accelerates indefinitely, i.e. $v_w \rightarrow 1$.
 \end{itemize}
In practice, we distinguish between the various regimes using $\eta$ as a classifier, with smaller values leading to runaway, then detonation, and finally larger values to hybrids or deflagrations.
We summarise the relationships between these velocities and the different regimes, in the wall frame, in Table \ref{tab:bubble}. Note that one has, for all regimes, $ \langle v \rangle = (v_+  + v_-)/2  \lesssim v_w$ as assumed for our definition of $\eta$.

\begin{table}[t]
    \centering
    \renewcommand{\arraystretch}{1.4} 
    \begin{tabular}{@{} LCCCL @{}}
        \toprule
        \text{Regime} & v_w & v_+ & v_- & \text{Features of the expansion} \\
        \midrule
\text{Detonation} & v_w > c_s & v_+ = v_w  & v_- < v_w & \text{Supersonic: rarefaction wave behind wall} \\
\text{Deflagration} & v_w < c_s & v_+ > 0 & v_- = v_w & \text{Subsonic: shock wave in front of the wall} \\
\text{Hybrid} & v_w > c_s & c_s>v_+ > 0 & v_- = c_s & \text{Sonic: rarefaction and shock wave}\\
\text{Runaway} & v_w \to 1 & v_+ \approx 0 & v_- \approx 0 & \text{No steady-state, walls accelerate indefinitely} \\
\bottomrule
    \end{tabular}
    \caption{\textit{Summary of the relationship between the bubble wall velocity, $v_w$, and the plasma velocities immediately in front of ($v_+$) and behind ($v_-$) the wall, for the distinct hydrodynamical regimes of bubble expansion. Throughout, the speed of sound in the relativistic plasma is taken to be $c_s = 1/\sqrt{3}$. See Appendix \ref{app:wallspeed} for details. \label{tab:bubble}}}
\end{table}

In the  model adopted in this work, for tiny values of the friction, the boundary between the different regimes becomes increasingly narrow so that numerically only the ``runaway'' and ``no-FOPT'' setup will be in practice visible. Note that the region accessible for deflagration and hybrid is actually slightly larger in the so-called \enquote{Local Thermodynamics Equilibrium} which neglects all out-of-equilibrium effects -- including the friction parameter \cite{Ai:2021kak}. Altogether, we found that using the latter only shifts the accessible region in terms of the flavour gauge coupling by a few percent, and so we opt to keep only the friction approach in our numerical results.

\par For the deflagration and the hybrid regimes, it is clear from the above description that we cannot solve the system from the matching equations and need to solve the full relativistic Euler equation derived from Eq.~\eqref{eq:TmunuconserveEq} instead.
This is typically done by assuming a spherically-symmetric configuration, considering $r$ as the distance from the centre of the bubble and $t$ as the time since nucleation. The variable $\xi = r/t$ is therefore the velocity of a given point in the wave profile; particles at this point move with velocity $v=v(\xi)$, with the wall velocity $v_w $ central to  the GW spectrum prediction.

In Appendix \ref{app:wallspeed}, we elaborate on the resultant system of differential equations required to solve for the wall velocity in each regime and the sampling algorithm that we employ to do so.

\begin{figure}[t]
    \centering
    \includegraphics[width=0.75\linewidth]{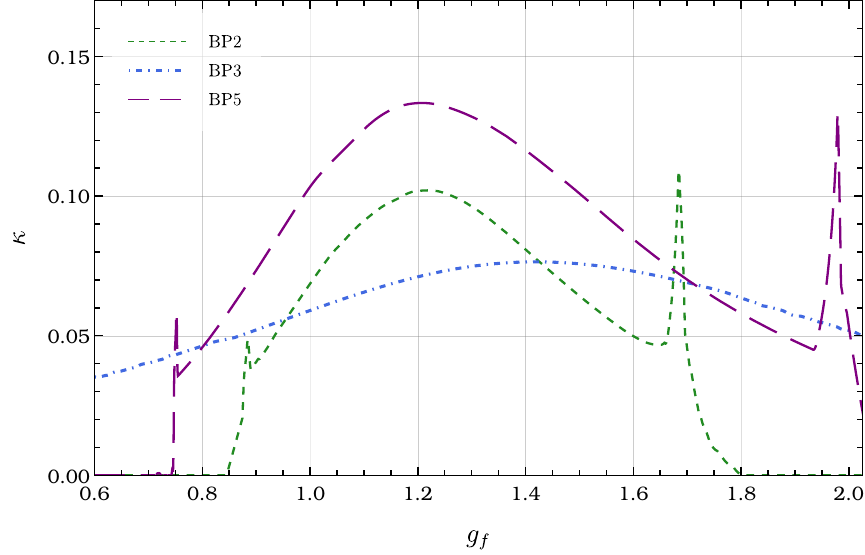}
    \includegraphics[width=0.495\linewidth]{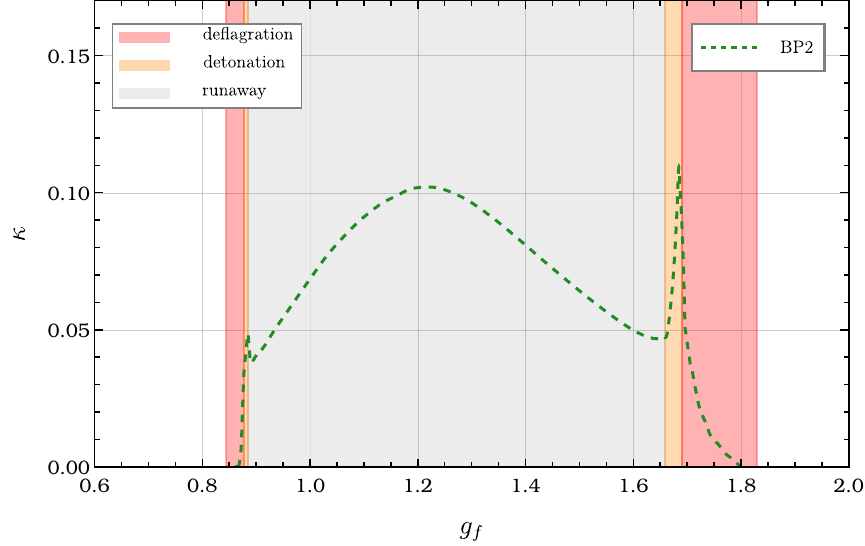}
    \includegraphics[width=0.495\linewidth]{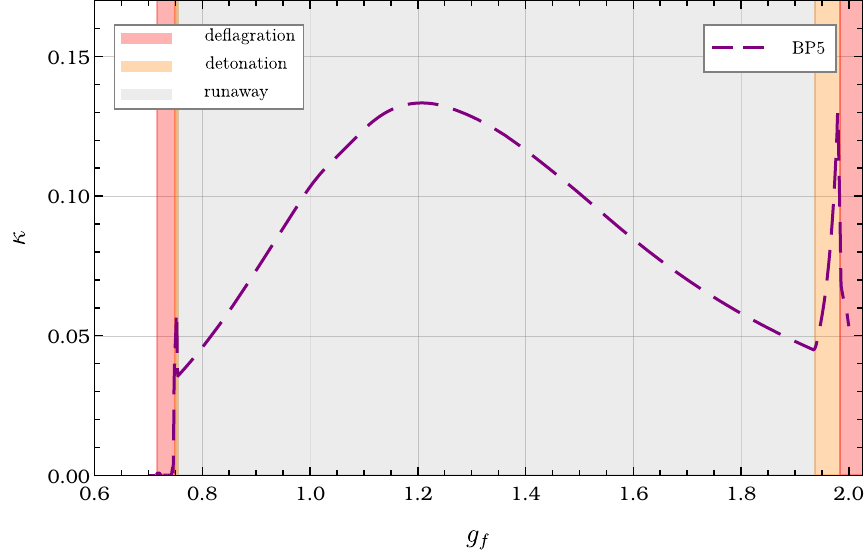}
    \caption{\textit{The evolution of $\kappa$, the efficiency parameter, with respect to $g_f$ for three benchmark points. Note the peaked behaviour as we transition from one regime to another, highlighted in the lower panels for BP2 (left) and BP5 (right). To guide the reader's eye in these plots, we connect across this discontinuous jump from the detonation to deflagration regimes. }}
    \label{fig:gfKappa}
\end{figure}

\paragraph{GW production and numerical results}
\par With the plasma thermodynamics obtained and the bubble wall velocity $v_w$ defined, we are able to express the extent to which the available energy of the phase transition (i.e. the latent heat, $\alpha_N \rho_R$) is converted into plasma kinetic energy: through the efficiency factor $\kappa$, defined in Ref. \cite{EllisLewicki:2018mja} as
\begin{equation} \label{eq:kappa}
    \kappa = \frac{3}{\alpha_N \rho_R v^3_w} \int^{v_w}_{c_s} w \, \xi^2 \frac{v^2}{1-v^2} \; d \xi \;.
\end{equation}
\noindent 
For the case of the runaway wall, we have directly the semi-analytical result~\cite{Espinosa:2010_EnergyBudget}:
\begin{equation} \label{eq:kappaRunaway}
    \kappa = \frac{\alpha_N}{0.73 + 0.083 \sqrt{\alpha_N} + \alpha_N}\;.
\end{equation}
Combining all the above results, we could then explore the behaviour of the phase transition bubbles for the BPs with a strong-enough FOPT: BP2, BP3, and BP5. We show in Fig.~\ref{fig:gfKappa} the efficiency factor as a function of the flavour gauge coupling $g_f$. Several features are noteworthy in these figures. First, one can notice that the range where $\kappa$ is non-zero is smaller than the one where a FOPT was predicted in Sec.~\ref{subsec:PTobs}. This is the direct consequence of the fact that the possibility to nucleate a bubble of a true vacuum does not guarantee that this bubble can actually grow.  Indeed, the driving force behind the bubble expansion is the imbalance between the pressure on both sides of the wall, which is at most equal to the difference between the pressure in the broken and symmetric phases, which is shown in Eq.~\eqref{eq:pminus} and Eq.~\eqref{eq:pplus}, respectively. This implies that we must at least have~\cite{Espinosa:2010_EnergyBudget}
\begin{align}
    \alpha_N ~\gtrsim~ \frac{1}{3} (1 - a_- / a_+ ) \ .
\end{align}
The second feature is that the efficiency peaks on the edge of our plot. As we show in the lower panel of Fig.~\ref{fig:gfKappa}, this corresponds to the system crossing from runaway bubbles toward detonation and deflagration steady-state solutions. Due to the low values of the friction parameters generated by the flavour gauge boson, such solutions exist only in a very limited range. Note that while a similar behaviour exists for low $g_f$ value, the tuning there is even more severe since the friction is proportional to $g_f^2$ via $m_{\rm D}^2$ so that we do not numerically observe these solutions. Note that when switching from the detonation to the deflagration regime, the wall parameters are not continuous. This is a consequence of choosing to retain only detonation solutions when both deflagration and detonation are possible steady-state solutions.
Finally, we never have hybrid solutions in our parameter space, in line with previous results~\cite{Espinosa:2010_EnergyBudget} demonstrating that for low friction, hybrid solutions are superseded by either detonation or runaway regime.

\par We can now focus on the predicted GW spectrum. According to the sound-shell model \cite{Hindmarsh:2015qta,Hindmarsh:2017_AcousticGWshape}, the acoustic perturbations (\enquote{sound shells}) around expanding bubble walls have a finite thickness that scales with the bubble size (or mean bubble separation); the sound-shell width sets the characteristic wave-number at which the acoustic source peaks. GWs are sourced by the fluid shear stresses produced where these sound shells overlap and where velocity and pressure gradients are large. These acoustic shear stresses are the dominant contribution to the GW spectrum for weak-to-intermediate transitions, $\alpha \sim 10^{-2}-10^{-1}$.

\par Following the numerical simulations performed in Refs~\cite{Hindmarsh:2015qta,Hindmarsh:2017_AcousticGWshape,
Guo:2020grp_Upsilon}, we can write the sound-wave contribution to the GW signal as summarised in Refs~\cite{Caprini:2019_GW-PT-LISA,Athron2023_GWreview},
\begin{equation}
\label{eq:GWOmega0}
    h^2 \Omega_{\text{sw}} (f) =  2.59 
    \times 10^{-6} 
    \left[ \left( \frac{g_*}{100} \right)^{-1/3} \right] \left( \frac{\kappa_{\text{sw}} \alpha}{1 + \alpha} \right) ^2 \left( \frac{\beta}{H_N} \right)^{-1} \text{max}(v_w, c_s) \Upsilon(\tau_{\text{sw}})
    S_{\text{sw}} \left( f, v_w \right)\;,
\end{equation}
\begin{equation}
\label{eq:GWf0}
    f^{\rm peak}_{\text{sw}} =
      8.9\times 10^{-6} \text{ Hz} \left[ \left(\frac{g_*}{100} \right)^{1/6}
      \left(\frac{T_N}{100 \text{ GeV}} \right) \right] 
      \frac{1}{ \text{max}(v_w, c_s) } \left(\frac{\beta}{H_N} \right) \left( \frac{z_p}{10} \right)\;.
\end{equation}
\noindent  
Here, $\kappa = \kappa_{\text{sw}}$, as defined in Eq. (\ref{eq:kappa}), and denotes the fraction of energy available for GWs sourced specifically from sound waves. In this case, the numerical prefactors absorb the contribution of the redshift factors enclosed in square brackets \cite{Athron2023_GWreview}. We set the simulation parameter $z_p$ to 10 \cite{Hindmarsh:2017_AcousticGWshape}. The spectral shape of the sound-wave GW contribution is well approximated by
\begin{equation}
    S_{\text{sw}} (f)  = \left( \frac{f}{f_{\text{sw}}^{peak}} \right)^3 \left(\frac{7}{4+3(f/f_{\text{sw}})^2}\right)^{7/2} \;.
\end{equation}
\noindent The lifetime suppression factor, $\Upsilon (\tau_{\text{sw}}) $, allows us to take into account the finite nature of the sound waves' lifetime. Within the radiation-dominated era,
\begin{equation}
    \Upsilon (\tau_{\text{sw}}) = 1 - \frac{1}{\sqrt{2 \tau_{\text{sw}} H_*}} \;.
\end{equation}
\noindent Here, we consider the asymptotic value, $\tau_{\text{sw}} H_N \rightarrow \infty$, such that $\Upsilon (\tau_{\text{sw}}) \rightarrow 1$ \cite{Guo:2020grp_Upsilon}. In so doing, we explicitly assume a very long lifetime of the sound waves and thereby suppress the influence of  non-linear shocks and turbulence. \\

\par Note that for wall velocities approaching the speed of light, the standard sound-wave template of Eq. \eqref{eq:GWf0} may overestimate the GW amplitude by a factor of up to an order of magnitude. This is due to enhanced shock formation, non-linear fluid behaviour, and resultant decreased efficiency in transferring vacuum energy into bulk fluid motion; the acoustic source lifetime is thereby reduced. The peak-frequency scaling, however, is unaffected \cite{Hindmarsh:2013xza,Hindmarsh:2015qta}. Upon comparing against signals predicted by the collision spectra summarised in Refs.~\cite{Caprini:2015zlo, Weir2017_GWsFOPTreview}, however, we found that the predicted peak amplitude of the bubble-collision contribution remained subdominant for all parameters studied in this work. After introducing the spectral shape of Ref.~\cite{Caprini:2015zlo}, the collision-generated GW signal remained subdominant to that of the sound waves across the frequency interval examined, $f \in [10^{-6},10^6]$.

\par This may be due to the fact that our transitions satisfy $\alpha < 1$; as stated earlier, sound-waves are expected to dominate as this is the regime in which the efficiency factor $\kappa_{\rm sw}$ capturing the transmission of energy from bubble wall into into bulk fluid motion remains sizeable \cite{Hindmarsh:2015qta,Hindmarsh:2017_AcousticGWshape}. As a result, the acoustic period is sufficiently long-lived and its GW contribution remains significant; the collision signal only becomes competitive for very strong, vacuum-dominated transitions. Altogether, these features ensure that the sound-wave spectra remained the leading contribution throughout our parameter space.

\subsection{Predictions from the $\sux$ model}
\begin{figure}[t!]
\centering
\includegraphics[width=0.495\linewidth]{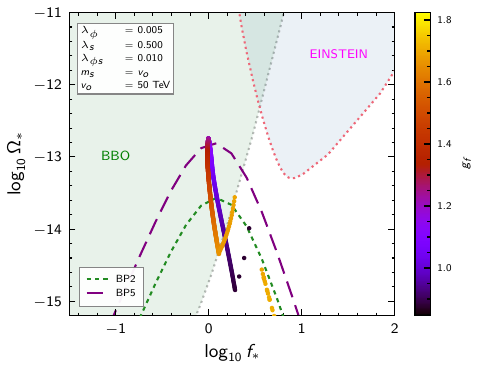}
\includegraphics[width=0.495\linewidth]{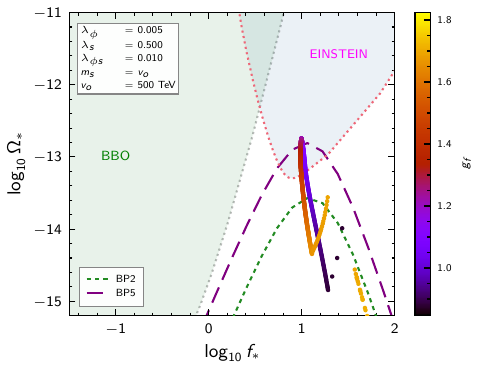}
\caption{\textit{Peak amplitudes vs peak frequencies of GW predictions for $v_0=50$ TeV (left) and $v_0 = 500$ TeV  (right), setting $g_f \in (0.6,2)$ for $\lambda_{\phi}=0.005$, $\lambda_s = 0.5$, and $\lambda_{\phi s} = 0.01$ (BP2, varying $g_f$); we include the explicit spectra of GWs for BP2 and BP5.}  }
\label{fig:GWforBP2}
\end{figure}
\begin{figure}[t]
\centering
\includegraphics[width=0.495\linewidth]{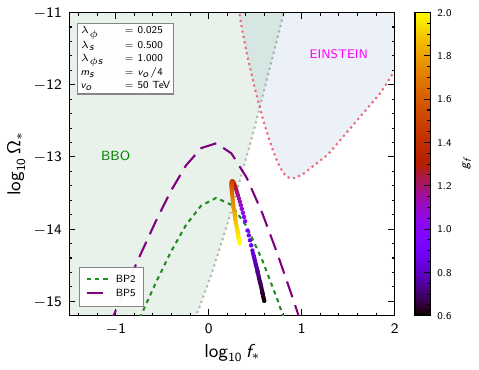}
\includegraphics[width=0.495\linewidth]{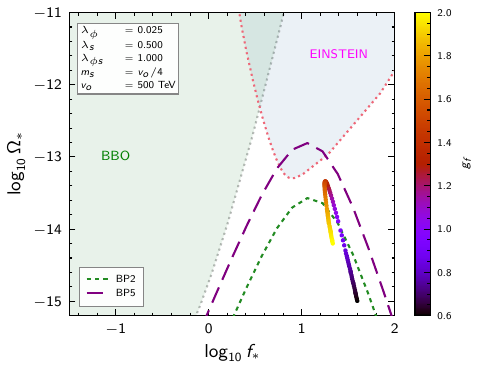}
\caption{\textit{Peak amplitudes vs peak frequencies of GW predictions for $v_0=50$ TeV (left) and $v_0 = 500$ TeV  (right), setting $g_f \in (0.6,2)$ for $\lambda_{\phi}=0.025$, $\lambda_s = 0.5$, and $\lambda_{\phi s} = 1.0$ (BP3, varying $g_f$); we include the explicit spectra of GWs for BP2 and BP5.} }
\label{fig:GWforBP3}
\end{figure}
\begin{figure}[t]
\centering
\includegraphics[width=0.495\linewidth]{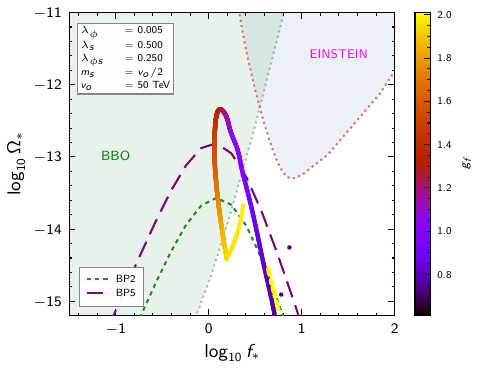}
\includegraphics[width=0.495\linewidth]{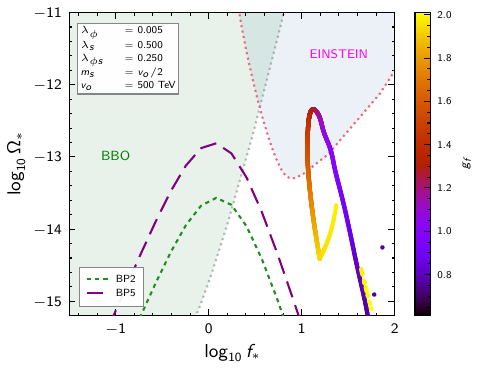}
\caption{\textit{Peak amplitudes vs peak frequencies of GW predictions for $g_f \in (0.6,2)$ for $v_0=50$ TeV (left) and $v_0 = 500$ TeV  (right), setting $\lambda_{\phi}=0.005$, $\lambda_s = 0.5$, and $\lambda_{\phi s} = 0.25$ (BP5, varying $g_f$); we include the explicit spectra of GWs for BP2 and BP5.} }
\label{fig:GWforBP5}
\end{figure}

We finally combine the tools developed in the previous section to project the sensitivity of various present and future GW observatories to the flavour phase transition in the horizontal gauge symmetry model of Sec.~\ref{sec:su2}.
We show in Figs~\ref{fig:GWforBP2}-\ref{fig:GWforBP5} the spectrum of stochastic GW for our benchmark points BP2, BP3, BP5 at present time, as estimated from Eqs~(\ref{eq:GWOmega0}) and~(\ref{eq:GWf0}), overlaid with the SNR estimate of the sensitivity from BBO and Einstein Telescope~\cite{Harry:2006fi,Punturo:2010zz}. In all instances we have taken as a typical flavour scale $v_0 = 500 \,\textrm{TeV}$, comfortably above typical flavour constraints, and $v_0 = 50 \,\textrm{TeV}$, close to the minimum value allowed by flavour physics constraints for our model. Note that the only effect of this shift is to change by a factor of $10$ the peak frequency, as expected. At the scale $v_0 = 500$ TeV,  BP5 and BP2 reach Einstein Telescope sensitivity, while BP5, BP3 and BP2 could be detectable in BBO for $v_0 = 50 \,\textrm{TeV}$. The result depends strongly on flavour gauge couplings, as shown in the coloured points in $v_0 = 50 \,\textrm{TeV}$ Figs.~\ref{fig:GWforBP5},~\ref{fig:GWforBP3} and ~\ref{fig:GWforBP2}. The latter show the frequency $f_{\text{sw}}^{peak}$ and amplitude $\Omega_{\text{sw}}^{peak}$ of the peak of the stochastic GW spectrum  (which is a strong marker of the overall detectability of the rest of the spectrum). For BP2, BP3, and BP5, we see that a significant range of flavour gauge couplings would lead to a detectable flavour phase transition. The evolution of $f_{\text{sw}}^{peak}$ and $\Omega_{\text{sw}}^{peak}$ is remarkably complex and reflects on the one hand the variation of the phase transition strength with $g_f$ presented in Fig.~\ref{fig:alphalm0-025} and on the other hand the impact of the bubble hydrodynamics via the efficiency parameter $\kappa$ shown in Fig.~\ref{fig:gfKappa}. In particular, we observe that the GW spectrum amplitude increases sharply once in the detonation regime, then discontinuously decreases to much lower value once our numerical calculation switches to the deflagration regime.

\begin{figure}[t]
\centering
\includegraphics[width=0.6\linewidth]{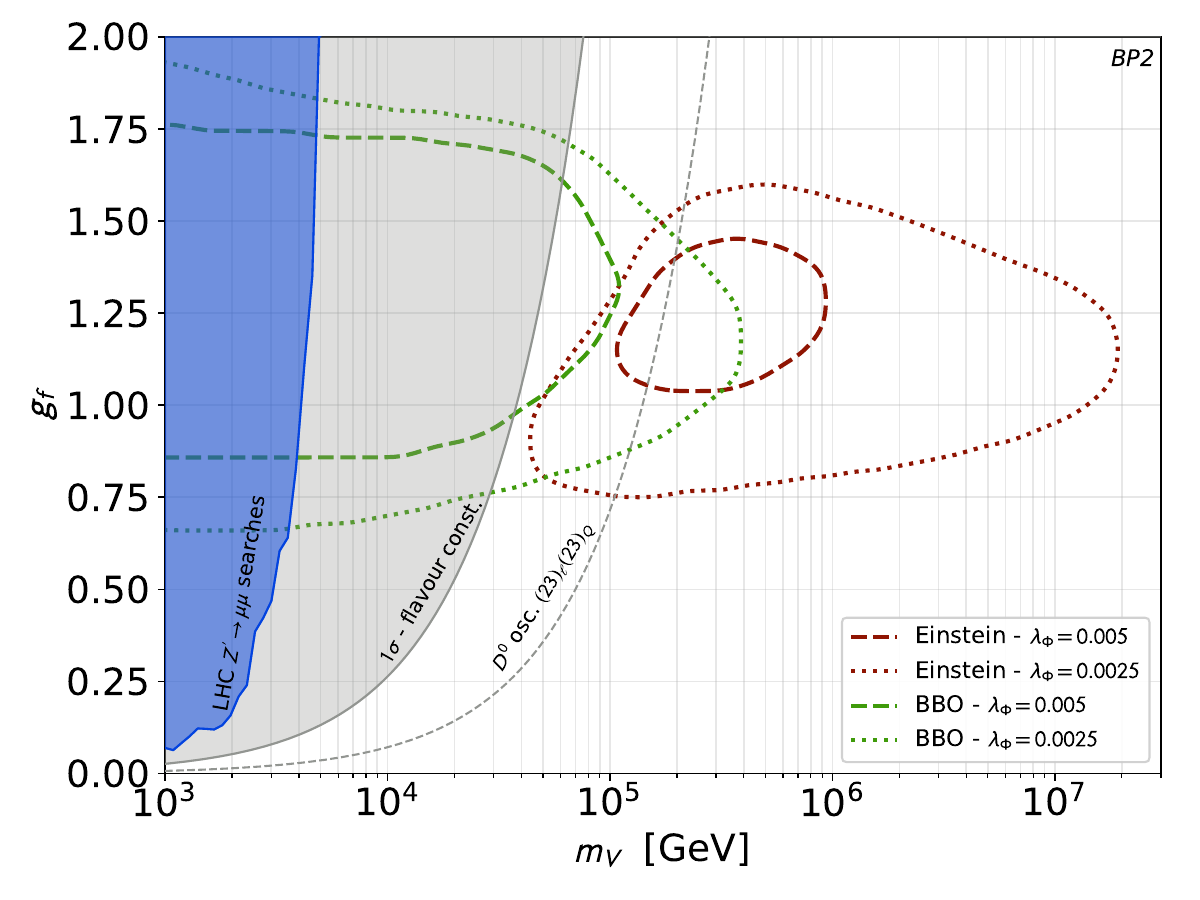}
\caption{Projected limits for BP1 as function of the gauge boson mass $m_V$ for $\lambda_\phi = 0.005$ (dashed line) and $\lambda_\phi = 0.0025$ (dotted line) for both Einstein Telescope (rust lines) and BBO (dark green lines).
}
\label{fig:ProjectedLimitBP2}
\end{figure}

\begin{figure}[t]
\centering
\includegraphics[width=0.49\linewidth]{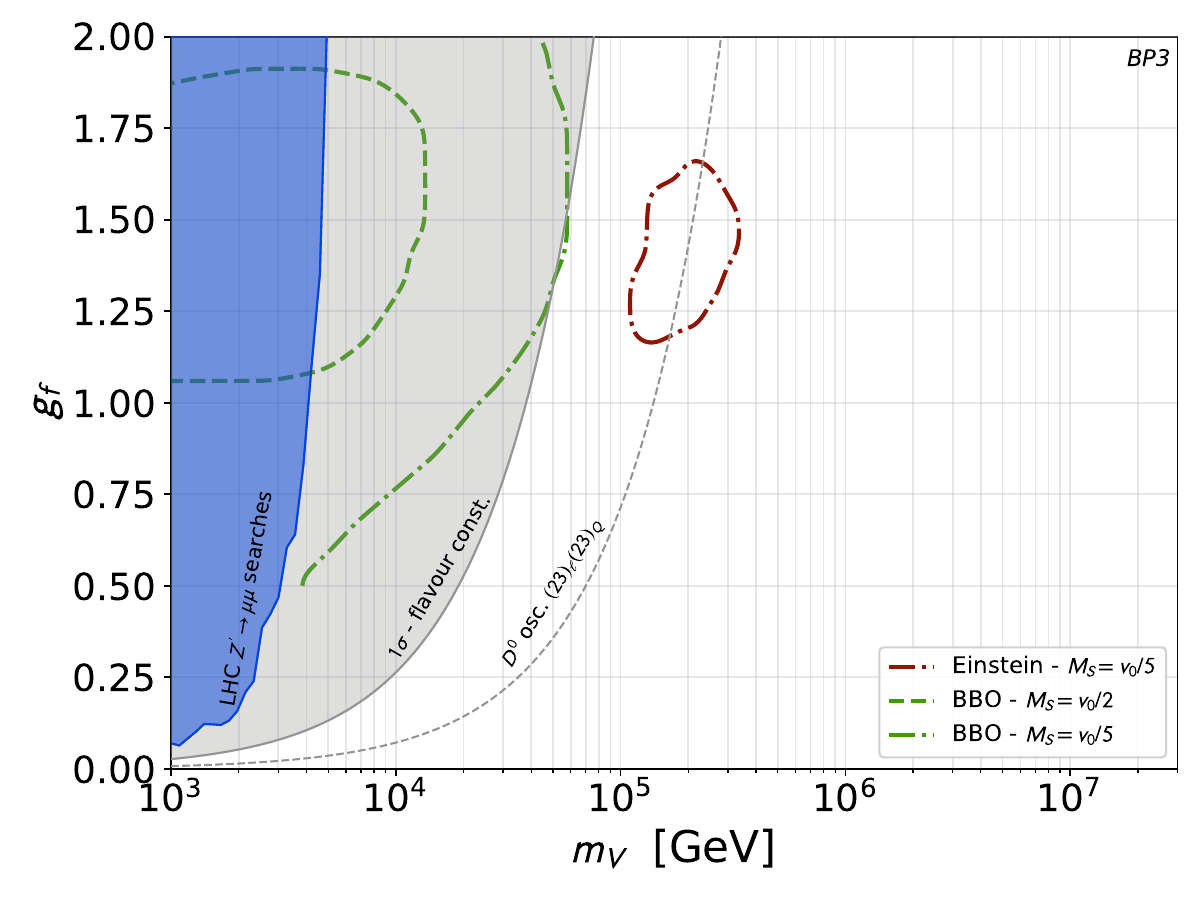}
\includegraphics[width=0.49\linewidth]{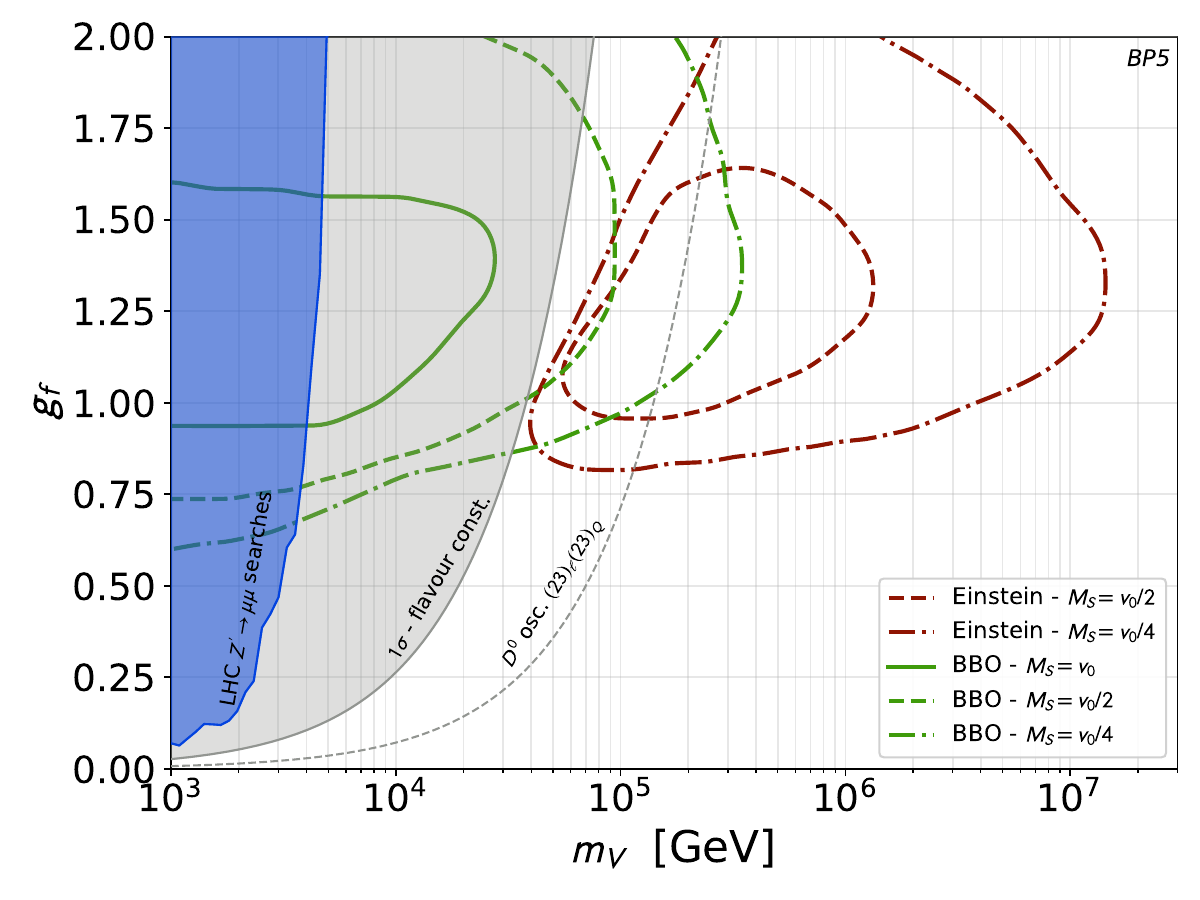}
\caption{Projected limits for BP3 and BP5 as function of the gauge boson mass $m_V$ for $m_S = v_0$ (full line), $m_S = v_0/2$ (dashed line) and $m_S = v_0/4$ (dotted line) for both Einstein Telescope (rust lines) and BBO (dark green lines).}
\label{fig:ProjectedLimitBP5}
\end{figure}

It is striking that the above detectability result depends strongly on the flavour gauge parameters and much more weakly on the actual flavour scale (or flavour pattern for the SM fermionic sector). We show the phenomenologically interesting result in Fig.~\ref{fig:ProjectedLimitBP2} for BP2. The red dashed (dotted) line represents the Einstein Telescope sensitivity for $\lambda_\phi = 0.005$ ($\lambda_\phi = 0.0025$), while green dashed (dotted) line the corresponding BBO sensitivity. The grey area represents the most conservative exclusion from flavour constraints as estimated in~\cite{DarmeDeandrea2023_SU2}. We further overlaid in dashed grey the limits $D^0$-meson oscillations discussed in Sec.~\ref{sec:su2} as an example of an individual flavour constraint for a specific flavour model. The blue region is the best current LHC constraint on this type of NP scenario~\cite{DarmeDeandrea2023_SU2,CMS:2019buh,CMS:2021ctt,CMS:2022fsw}. With limits extending up to $\sim 20 \textrm{ PeV}$, this illustrates our main conclusion: for order one gauge coupling, Einstein Telescope could probe relevant flavour scenarios up to scales more than two orders of magnitude higher than current flavour constraints and nearly four orders of magnitude larger than current LHC searches. While BBO could detect fainter phase transitions, this translates into a marginal improvement in flavour gauge couplings as the phase transitions very quickly become second order. Additionally, for the flavour models considered in this study, the BBO detectability range tends to overlap with the parameter space already tested via flavour processes.
Furthermore, while obtaining a strong enough phase transition requires a small quartic coupling $\lambda_\phi$ in the absence of LQs, we stress that this requirement is strongly relaxed when we include these additional degrees of freedom. We illustrate this point in Fig.~\ref{fig:ProjectedLimitBP5} for BP3 (left) and BP5 (right)  by showing the projected reach of Einstein Telescope and BBO for various LQ masses. 

We see that the presence of these new degrees of freedom around the phase transition scale supports strong FOPT up to quartic couplings at the $\mathcal{O}(0.01)$ level.  
The presence of additional degrees of freedom (such as new VLFs, extra LQs, etc.) is a staple of models explaining the SM flavour structure using new gauge groups and we thus expect that strong FOPT is a generic feature of these theories rather than an exception. As a further check, we observe that tripling the number of LQ degrees of freedom (for instance, with a larger gauge group representation) leads to a further weakening of the constraint of the quartic coupling to the $\mathcal{O}(0.05)$ level, remarkably close to the SM value at these energy scales. Thus, while small quartic couplings seem always required for the presence of a GW signal, the requirement will not likely be as stringent in complete flavour models with their numerous new VLF or scalar fields than in simplified models with only a new $SU(2)$ gauge group. Note that the lower mass on the gauge coupling itself is only mildly affected and hardly extends by $\sim 0.7$ in all of our benchmark points.

\newpage
\section{Conclusion}
\label{sec:conclusion}

We have shown that models of flavour based on $SU(2)$ horizontal gauge symmetry can undergo a FOPT for relatively generic scalar potential structures, due in large part to the expected presence of additional degrees of freedom around the phase transition scales. Consequently, these models will lead to a detectable GW signal in future observatories for an extremely wide range of energy scales, up to $\sim 10^7$ GeV. The main limitation of this reach is the requirement that the flavour gauge coupling is of order $\mathcal{O}(1)$, with couplings below $g_f \lesssim 0.7$ difficult to probe as they typically do not lead to a FOPT. 

With the flavour constraints that we have reviewed, pushing the typical scale of the breaking of the horizontal gauge flavour symmetry above $\sim 40$ TeV, it is clear that the flavour gauge bosons themselves will hardly be accessible at colliders in the foreseeable future, particularly if the new gauge coupling is sizeable (and thus the gauge boson masses are of the order of the flavour-breaking VEV). We have thus calculated the thermal effective potential in this regime using state-of-the-art dimensional reduction techniques based on the \texttt{DRalgo} code, and verified their agreement with the standard resumed Truncated Full Dressing approach. We found that the phase transition strength reaches an optimum for $g_f \sim 1-1.5$ and does not benefit from a further increase due to loop corrections on the quartic interaction eventually overcoming the tree-level contribution. We obtained the tunnelling rates using the \texttt{FindBounce} numerical routines and leveraged the corresponding nucleation temperature to extract the typical strength and duration of the possible phase transitions.    

We found that strong FOPTs can typically be achieved for quartic interactions in the $\mathcal{O}(0.005)$ region in the absence of additional particles in the theory. However, new particles -- typically VLFs or LQs -- are usually required by complete flavour theories, for the generation of the flavour hierarchies in the Yukawa matrices. We have shown that a single LQ with mass around the flavour-breaking VEV is enough to allow strong FOPTs for quartic interactions in the $\mathcal{O}(0.025)$ range. With complete flavour models typically introducing several such objects or many VLFs, we thus expect that SM-size quartic couplings would be enough to generate a FOPT in typical complete flavour models. 

The phase transitions themselves were described in an effective approach using an overall friction parameter $\eta$ to describe the out-of-equilibrium effects. The dominant contributions to this friction arose from the soft modes of the flavour gauge bosons themselves, whose interactions with the SM plasma were described via the Langevin equation, then matched to $\eta$. We numerically solved the hydrodynamics equations to identify the adequate regime for each phase transition, additionally retaining only runaway scenarios for strong FOPTs with strength larger than $\alpha \sim 0.1$ in light of recent field simulations pointing towards the non-thermalisation of the plasma shells in front of the bubble, preventing the occurrence of stationary solutions.

We finally presented the expected GW spectrum, focussing on the dominant sound-wave component. When increasing the flavour gauge coupling, the amplitude of the GW signal typically increased until it reached a maximum at order one values. While the phase transition strength then decreased due to the dominance of one-loop contributions into the thermal potential, the friction kept on increasing, leading to a part of our parameter space presenting  phase transitions with stationary solutions in the detonation and deflagration regimes, where the vacuum energy is very efficiently transmitted to the GW spectrum and locally increases the predicted spectrum's amplitude. 

In the parameter space that is not currently probed by flavour experiments, we found that the projected Einstein Telescope was the most relevant, with BBO only marginally improving flavour constraints. In all cases, the remarkable frequency range detection prospect of GW observatories means that these scenarios can be tested over several orders of magnitude, even while being limited to order one flavour gauge couplings, thereby testing a parameter space highly complementary to flavour or collider searches.

\acknowledgments
AC acknowledges the support of the Initiative Physique des Infinis (IPI), a research training programmme of Idex SUPER at Sorbonne Universit\'e.
AD and LD thank Sebastian Monath and Florian Nortier for many insightful discussions during the course of this work.
ASC was partly supported by the National Research Foundation of South Africa, and wishes to thank New York University - Abu Dhabi for their  hospitality during the completion of this work.
\paragraph*{\textit{Software}} As stated in the text, the finite-temperature effective potential and the bubble action were computed within \texttt{Wolfram Mathematica}~\cite{mathematica}: dimensional reduction was facilitated by the \texttt{DRalgo} package~\cite{DRalgo}, which itself relies on \texttt{GroupMath}~\cite{GroupMath}, and field bounce solutions were obtained using \texttt{FindBounce}. Numerical solving of the hydrodynamic equations was carried out using the \texttt{NumPy}~\cite{numpy}, \texttt{SymPy}~\cite{SymPy} and \texttt{SciPy}~\cite{scipy} libraries, with sampling performed using \texttt{Dynesty} \cite{dynesty}. The plots were produced with \texttt{Matplotlib}~\cite{matplotlib}, together with the \texttt{SciencePlots} \cite{SciencePlots} style package.

\newpage
\begin{appendices}

\section{Thermal and hydrodynamic parameters for the benchmark points \label{app:BPparams}}

The parameters sensitive to the scale $v_0$ are restricted to $T_c, T_N, \phi_c, \phi_N,$ and $f_N$. They increase by an order of magnitude as we scale $v_0 = 500$ TeV. All benchmark points are expected to lead to a runaway regime during the phase transition, highlighting the fact that obtaining a steady-state solution requires some mild tuning on the original Lagrangian parameters.
\begin{table}[hb!]
    \centering
    \caption{\textit{Thermal and hydrodynamic parameters computed for each of the five benchmark points, where the $S$ LQ is kept at the US scale. We set $m_{\phi}^2 = \lambda_{\phi}v_0^2, \; m_s^2 =v_0^2, \;  \lambda_{s}=0.5,$ and $v_0=50$ TeV.}}
    \setlength\tabcolsep{0.1cm}
    \def\arraystretch{1.5}
    \begin{tabular}{@{}C|CCCCC@{}}
    \hline\noalign{\smallskip}
         & \text{BP1} & \text{BP2} & \text{BP3} & \text{BP4} & \text{BP5} \\ 
         \hline
        m_S/v_0 & 1 & 1 & 0.25 & 0.1 & 0.5 \\ 
        \lambda_{\phi} & 0.025 & 0.005 & 0.025 & 0.1 & 0.005 \\ 
        \lambda_{\phi S} & 0 & 0.01 & 1 & 1.5 & 0.25 \\ 
        g_f & 1.5 & 1 & 1 & 1.5 & 1.5 \\ 
        \hline
        T_c \; \rm{[TeV]} & 21.28 & 16.95 & 23.61 & 25.76 & 17.37 \\ 
        \phi_c \; \rm{[TeV]} & 46.13 & 72.30 & 78.28 & 56.38 & 67.83 \\ 
        T_N \; \rm{[TeV]} & 19.06 & 12.34 & 19.50 & 24.09 & 13.76 \\ 
        \phi_N \; \rm{[TeV]} & 51.78 & 88.71 & 87.95 & 60.71 & 80.60 \\ 
        \eta & 0.0022 & 0.0037 & 0.0013 & 0.0020 & 0.0078 \\ 
        \alpha_N & 0.011 & 0.055 & 0.048 & 0.021 & 0.085 \\ 
        \beta/H_N & 1886 & 1123 & 1339 & 2850 & 884 \\ 
        v_w & 0.5756 & 0.7877 & 0.6930 & 0.4870 & 0.7387 \\
        L_w \; \rm{[TeV]}^{-1}& 0.0507 & 0.0999 & 0.0514 & 0.0472 & 0.0687 \\ 
        v_J & 0.6561 & 0.7365 & 0.7281 & 0.6832 & 0.7649 \\
        \hline
        \alpha_+  & 0 & 0.054 & 0.048 & 0 &0.085 \\ 
         \xi_w  & 0 & 1 & 1 & 0 & 1 \\ 
         \kappa  & 0 & 0.0688 & 0.0607 & 0 & 0.1010 \\
         \text{regime} & - &  \text{runaway} &  \text{runaway} &  - &  \text{runaway} \\ 
        h^2 \Omega^{\rm peak}_{\text{sw}} & 0 & 2.76^{-14} & 1.40^{-14} & 0 & 1.67^{-13} \\ 
        f^{\rm peak}_{\text{sw}} \ \textrm{[Hz]} & 0 & 1.29 & 2.43 & 0 & 1.13 \\ 
\hline\noalign{\smallskip}
    \end{tabular}
    \label{tab:BPparamsComplete}
\end{table}

\newpage
\section{Truncated Full Dressing versus Dimensional Reduction
\label{app:4dtheory}}

\par As discussed in Sec. \ref{subsec:DRexplained}, the breakdown of the perturbative expansion at high temperatures is due to the infrared divergences. In a scalar theory, this is evidenced in the \enquote{daisy} or \enquote{ring} diagrams, where the zero infrared modes are screened by the non-zero ultraviolet modes that dominate at each loop order, such that the scalar self-coupling diverges. The traditional approach to solving this problem is referred to as \enquote{daisy resummation}: a technique through which the divergent diagrams are resummed, and a temperature-dependent mass counter-term is introduced to prevent double-counting. Two prescriptions exist to implement this thermal resummation. The first is the \enquote{Parwani} \cite{Parwani1991_Method}/\enquote{Truncated Full Dressing} (TFD) \cite{Curtin:2016_ThermalResum} approach, which directly replaces the scalar and longitudinal gauge boson masses within the potential with their resummed versions: $m^2_i(\phi)\to m^2_i(\phi)+\Pi_i(T)$, where $\Pi_i(T)$ is the Debye mass.\footnote{In Ref. \cite{Curtin:2016_ThermalResum}, the authors refer to the direct substitution of thermal masses into the effective potential as the \enquote{Full Dressing} approach. Since the thermal mass $\Pi$ is explicitly evaluated only to leading order in the high-temperature expansion, the method is quantified as the \enquote{Truncated Full Dressing} approach.} The \enquote{Arnold-Espinosa} \cite{ArnoldEspinoa1992_Method} method, introduces a daisy correction term into the effective potential,
\begin{align}
  V_\text{daisy}(\phi,T)=-\frac{T}{12\pi}\sum_i n_i \left[\left(m^2(\phi)+\Pi(T)\right)_i^{3/2}-\left(m^2(\phi)\right)_i^{3/2}\right] \, ,
\end{align}
where $n_i$ is the number of degrees of freedom for each species. The first term implements the resummation, while the second correcting removes the double counting.

\par Both methods have limitations. The Arnold-Espinosa approach resums only the zero Matsubara modes and so risks undercounting higher-order corrections. The Parwani/TFD approach, on the other hand, introduces thermal corrections directly into each mass term within the loop-level potential, effectively resumming all Matsubara modes $-$ but possibly over-counting as a consequence. Beyond these, both thermal resummation strategies are plagued by well-documented theoretical uncertainties \cite{Croon:2020_ThermalResum}, and therefore valid only in the high-temperature limit. 

\par In certain cases, however, the result of the TFD approach may contain a lesser scale dependence \cite{Curtin2024_UpdateOnOPD}. Recall from Table 3 of Ref. \cite{Croon:2020_ThermalResum} that this scale dependence introduces the largest contribution of theoretical uncertainty into the 4d approach to thermal resummation.  
\par For this reason, we favour the TFD technique in this appendix, and validate our results by illustrating its agreement with the DR approach for a simple manifestation of the  $\sux$ model studied in this work.

\par Let us begin with the tree-level potential after symmetry breaking, 
\begin{equation}\label{eqn:V0}
V_{\textrm{tree}}(\phi)=-\frac{1}{2}\mu^2_{\phi} \phi^2 +\frac{1}{4}\lambda_{\phi} \phi^4 + \frac{1}{2} \mu^2_s \vert s \vert^2 + \frac{1}{4} \lambda_s \vert s \vert^4 + \frac{1}{2} \lambda_{\phi s} \phi^2 \vert s \vert^2\;.
\end{equation}
For simplicity, we shall assume here that $\phi$ weakly couples to the complex scalar LQ, $\lambda_{\phi s} \sim 0$. Following the Minimal Subtraction $\overline{MS}$ renormalisation scheme, the loop-level zero-temperature correction is given by the Coleman-Weinberg potential,
\begin{equation}\label{eqn:V1}
 V_{1-\rm loop}(\phi)=\sum_{i= \phi,\chi,f,s}
  \pm  \frac{n_i}{64 \pi^2} m^4_i \left[ \log \left( \frac{m^2_i}{\mu^2} \right) - C_i\right] \;,
\end{equation}
\noindent where $\mu$ is the renormalisation scale and $C_i = 5/6$ $(3/2)$ for gauge bosons (scalars/fermions). Here, $+$ $(-)$ is used for bosons (fermions), and $m_i=m_i(v_{\phi})$, which is the tree-level mass calculated at the VEV $\langle s \rangle =0, \ \langle \phi \rangle =v_{\phi}$. The number of degrees of freedom $n$ for each species is summarised in Table \ref{tab:model}, where the hierarchy of scales $m_{\rm VLF} > m_{R\rm _u}, m_{ R_d} \gg m_s > v_{\phi} \gg v_{\rm EW}$ suggests that heavy species decouple from the thermal bath and therefore play a negligible role in the dynamics of the phase transition. As such, the degrees of freedom and the masses of the contributing species are, respectively, $ n_{\{\phi,\chi,s,f\}}=\{1,3,12,9\}$ and 
$m^2_{\{\phi,\chi,s,f\}}~=~\{ 3\lambda_{\phi} \phi^2 - \mu^2_{\phi},\lambda_{\phi}\phi^2 - \mu^2_{\phi},\lambda_s \phi^2/2 + \mu^2_s,g^2_f \phi^2/4 \}$, with the Goldstone degrees of freedom being $\chi$. 
\par For bosons (B) and fermions (F), the loop-level finite temperature corrections are 
\begin{equation}\label{eq:VT}
 V_T(\phi,T)=\sum_{i } \frac{n_iT^4}{2\pi^2} J_B\left(\frac{m^2_i}{T^2}\right)+\sum_{i }\frac{n_iT^4}{2\pi^2} J_F\left(\frac{m^2_i}{T^2}\right) \;,
 \end{equation}
\begin{equation}
J_{B/F }\left(a\right)= \pm\int_o^\infty dy y^2\log \left[1\mp e^{-\sqrt{y^2+a}}\right] 
\end{equation}
\noindent for $a = m^2_i/T^2$ \cite{Curtin:2016_ThermalResum,Huang2020_PatiSalamDynamics}. Within the high-temperature regime ($a =m^2_i/T^2\ll1$), we can expand the thermal integral $J_{B/F }$ as 
\begin{equation}
\begin{split}
J_B^{high}\left(a\right)& \approx -\frac{\pi^4}{45}+\frac{\pi^2}{12}a-\frac{\pi}{6}a^{3/2}-\frac{a^2}{32}\left(\log \left(a\right) -c_B\right) \;,\\
J_F^{high}\left(a \right)& \approx-\frac{7\pi^4}{360}+\frac{\pi^2}{24}a +\frac{a^2}{32}\left(\log\left(a\right)-c_F\right)\,,\label{eq:high}
\end{split}
\end{equation}
\noindent for $c_B=3/2-2\gamma_E+2\log\left(4\pi\right)$ and $c_F=3/2-2\gamma_E+2\log\left(\pi\right)$, with $\gamma_E\approx0.5772$. When $a =m^2_i/T^2 \gg1$, we consider the low-temperature approximation, 
\begin{equation}
J_{B,F}^{low} \left(a\right)  \approx - \sqrt{\frac{\pi}{2}}a^{3/4}e^{- \sqrt{a}}
\left(1 +  \frac{15}{8} a^{- 1/2} + \frac{105}{128} a^{-1}\right) \;. \label{eq:low}
\end{equation}

\noindent We connect the high and low temperature contributions using \cite{Huang2020_PatiSalamDynamics}
\begin{equation}
\begin{split}
J_B\left(a\right) & \approx e^{-\left(\frac{a}{6.3}\right)^4} J_B^{high}\left(a\right)+\left(1-e^{-\left(\frac{a}{6.3}\right)^4}\right) J_{B}^{low}\left(a\right) \;, \label{eq:lowconnect}\\
J_F \left(a\right) & \approx e^{-\left(\frac{a}{3.25}\right)^4} J_F^{high}\left(a\right) +\left(1-e^{-\left(\frac{a}{3.25}\right)^4}\right) J_{F}^{low}\left(a\right) \;.\\
\end{split}
\end{equation}
\noindent We substitute these into Eq. (\ref{eq:VT}) for each species $i= \phi,\chi,f,s$.

\par Finally, we must contend with the higher-order thermal corrections from scalars and longitudinal polarisations of the gauge bosons, with species $i$ in the centre of the \enquote{daisy diagram} and the relevant bosonic degrees of freedom $j$ in the outside rings. For the $\phi$ field, $\pi_{\rm{\phi}}$ must include the self-interaction term $\pi_{\rm{\phi}}^{\rm{\phi}}$ and the contribution from the gauge bosons $\pi_{\rm{\phi}}^{\rm{f}}$; the same applies to the Goldstone \cite{Carrington1991_FiniteTempSM,EllisLewicki:2018mja}. Similarly, the $\sux$ gauge bosons include the self-interaction term $\pi_{\rm{f}}^{\rm{f}}$, as well as contributions from the $\phi$ field $\pi_{\rm{f}}^{\rm{\phi}}$ and the SM fermionic degrees of freedom \cite{Huang2020_PatiSalamDynamics,EllisLewicki:2018mja}. Altogether, these are, respectively,
\begin{align}
    &\pi_{\rm{\phi}} = \pi_{\rm{\chi}} = \frac{\lambda_{\phi}}{2} + \frac{9}{48}g_f^2 \\
    & \pi_{\rm{f}} = \frac{2}{3} g_f^2 + \frac{1}{6}g_f^2 + \frac{1}{2} \left( 1 \times 2 + 3 \times 2 \right)g_f^2 = \frac{3 }{2} g_f^2 \;.
\end{align}

\begin{figure}[ht!]
\centering
\includegraphics[width=0.85\linewidth]{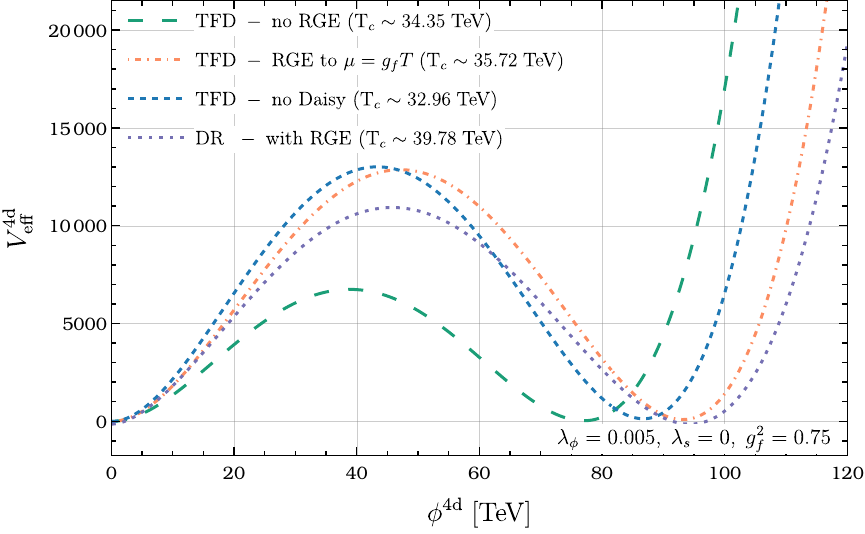}
\caption{\textit{For the benchmark point $\{\lambda_{\phi},g^2_f,\lambda_s \} = \{0.005,0.75,0 \}$ and $\mu_{\rm{ini}}=50$ TeV, we compare the effective potentials at $T=T_c$ calculated using the DR (\enquote{\texttt{DRalgo}}, indigo) and TFD approaches. For the latter, we observe the effect of including RG running to $\{\lambda_{\phi},g^2_f,\lambda_s \} = \{0.0045,0.7510,0 \}$ and $\mu_{\rm{fin}} \sim 30$ TeV (\enquote{with RGE}, salmon) and excluding it (\enquote{no RGE}, green), as well as the exclusion of the thermal mass corrections (\enquote{no Daisy}, blue).}}  
\label{fig:TFDvsDR}
\end{figure}

\noindent Recall that the contribution from the $\sux$ gauge boson interacting with the scalar field $\phi$ is governed by
\begin{equation}
    \pi^{j}_{scalar} (0) = \frac{1}{12} \frac{m^2_i(v_{\phi})}{v_{\phi}}T^2 \;,
\end{equation}
\noindent whereas the self-interacting term follows
\begin{equation}
    \pi^{L,V}_{SU(N)} = \frac{N}{3} T^2 \;.
\end{equation}
As such, the finite-temperature contribution to the potential is produced by evaluating Eq. (\ref{eq:VT}) at $m^2_{i} \rightarrow m^2_{i} + \sum_j \pi^j_i T^2$ with the approximations of Eq. (\ref{eq:lowconnect}). The full finite temperature effective potential is the sum of Eqs. (\ref{eqn:V0}), (\ref{eqn:V1}), and (\ref{eq:VT}), which we plot in Fig. \ref{fig:TFDvsDR}.

\par Recall from our discussions in Sec. \ref{subsec:DRimplementation} that to observe a FOPT, we find that the quartic of the potential must be small. In other words, the potential is very sensitive to the gauge-induced term, which is large and thus varies quite fast with the running. 
\par Due to this sensitivity of the potential to the gauge-induced term, we can in the 4d theory set $\mu_{\phi} = \mu_s =100$ TeV, and for the couplings $\{\lambda_{\phi} , g^2_f , \lambda_s \} = \{0.0075,0.75, 0 \}$. We set the RG-running from $\mu = 50$ TeV to $\mu \sim g_f T_c$ (approximately 30 TeV). For these values, we find degenerate minima for
\begin{align} \label{eq:VeffFOPT?}
\frac{v_{\phi}}{T_c} \Bigg \vert_{\rm{DR}} \approx 2.40 \;; \quad \quad \frac{v_{\phi}}{T_c} \Bigg \vert_{\rm{TFD}} \approx 
\begin{cases}
2.38\;, \quad \rm{ no \; RGE} \;;\\
2.60\;, \quad \rm{ RGE} \;;\\
2.63 \;,\quad \rm{ no \; Daisy}  \;.
\end{cases}
\end{align} 

\par Note that \enquote{RGE} indicates RG running \textit{and} thermal resummation. Since these satisfy the criterion $v_{\phi} / T_c > 1$ for a strong FOPT, we can expect a GW signal from this model that falls within the observational window. 

\par We plot the corresponding effective potential in Fig. \ref{fig:TFDvsDR}. The theoretical uncertainty carried by the TFD approach manifests itself in the form of a non-zero imaginary component of the effective potential, which has been related to the growth rate of long-wavelength modes around a constant background field \cite{WeinbergWu1987_ComplpexEffectivePotentials}. If this decay rate is not exponentially
suppressed, it can overcome bubble nucleation in certain contexts and lead away from the broken phase to a new phase with a non-homogeneous VEV \cite{Croon:2020_ThermalResum}. We treat the presence of this non-zero imaginary part as a source of systematic uncertainty  \cite{Croon:2020_ThermalResum}, with the assumption that such an error is permissible in the event that the imaginary component is much smaller than its real counterpart at the minima of the effective potential (as is the case here), following the standard practice in the literature \cite{DelaunayGrojean2007_EWSBnonrenormalisable}.

\par When comparing the curves in Fig. \ref{fig:TFDvsDR}, we focus on the height of the barrier and the value of the VEV. Recall that a high potential barrier is indicative of a strong FOPT. For the chosen benchmark point, we observe that the tallest barrier corresponds to the TFD approaches: in the absence of thermal corrections and then inclusive of the RG running. This indicates that resummation weakens the barrier, and the resulting phase transition; this is corroborated by the shift in $v_{\phi}/T_c$ from greater than 2.60 to 2.38 in Eq. (\ref{eq:VeffFOPT?}). We see an increase in the barrier height and in $v_{\phi}/T_c$ from 2.38 to 2.60 when thermal corrections are included with RG running. As such, thermal corrections should not be introduced without compensation from the RG running. In this scenario, we highlight that the running of the couplings decreases $\lambda_{\phi}$ and increases $g_f$: reducing the quartic term is known to improve the strength of the FOPT. Moreover, we observe that DR and TFD with thermal corrections and RG running share a VEV at their respective values of $T=T_c$. This indicates consistency between the DR and TFD approaches, provided that corrections are treated appropriately for the latter.

\newpage
\section{Calculation of the efficiency parameter and the wall velocity}
\label{app:wallspeed}
\renewcommand\thesubsubsection{\thesection.\arabic{subsubsection}}
Calculating the bubble wall velocity and efficiency factor $\kappa$ is essential to determine the GW spectrum, as $\kappa$ directly enters Eq.~(\ref{eq:kappa}). This appendix outlines the method employed to determine both $v_w$ and $\kappa$ across the different bubble expansion regimes summarised in Table~\ref{tab:bubble}. We explain the manner in which we apply a sampling algorithm in which to compute these parameters, where the inputs $ \{ \eta, \alpha_N, T_c, T_N, \beta/H_N \}$ are defined in Sec. \ref{subsec:PTobs} ($T_N$, $\beta/H_N$, and $\alpha_N$ in Eqs. (\ref{eq:Tn}), (\ref{eq:betaH}), and (\ref{eq:alphaN}), respectively). 

\subsubsection*{Profiles}
\label{app:profiles}

To compute $\kappa$, we first determine the enthalpy $w$, the fluid velocity $v=v(\xi)$ in the frame of the bubble centre, and the self-similar variable $\xi = r/t$. These quantities satisfy the conservation equations for a spherically symmetric relativistic fluid in self-similar coordinates~\cite{Espinosa:2010_EnergyBudget}:
\begin{equation}
\label{eq:edo}
\begin{cases}
\frac{dv}{d\tau} = 2vc_s^2(1 - v^2)(1 - \xi v), \\
\frac{d\xi}{d\tau} = \xi \left[ (\xi - v)^2 - c_s^2(1 - \xi v)^2 \right], \\
\frac{dw}{d\tau} = w \frac{1 + c_s^{-2}}{1 - v^2} \frac{\xi - v}{1 - \xi v} \frac{dv}{d\tau},
\end{cases}
\end{equation}
where $\tau$ is an auxiliary parameter and $c_s = 1/\sqrt{3}$ is the speed of sound, assumed constant across the bubble wall. Initial conditions are imposed at the bubble wall.

\subsubsection*{Determination of the regime}

The relationship between plasma velocities in front of ($v_+$) and behind ($v_-$) the wall follows from hydrodynamic matching~\cite{Espinosa:2010_EnergyBudget}:
\begin{equation}
\label{eq:vplus}
v_{+} = \frac{1}{1 + \alpha_{+}} \left[ \frac{v_{-}}{2} + \frac{1}{6v_{-}} \pm \sqrt{\left( \frac{v_{-}}{2} + \frac{1}{6v_{-}} \right)^2 + \alpha_{+}^2 + \frac{2}{3}\alpha_{+} - \frac{1}{3}} \right],
\end{equation}
where $\alpha_+ = \epsilon/(a_+ T_+^4)$ parameterises the strength of the phase transition in front of the bubble wall (c.f. Eqs. (\ref{eq:alphaN}) and (\ref{vplus})). The sign in Eq.~(\ref{eq:vplus}) determines the hydrodynamic regime: the plus sign corresponds to detonation, the minus sign to deflagration. Hybrid regimes interpolate between the two, while for runaway solutions, $v_w \rightarrow 1$.

\subsubsection*{Initial conditions}
\label{app:initialconditions}

The first moment of the planar-wall approximation of the scalar field equation of motion, Eqs.~(\ref{eq:EoMphiEspinosa}) and \eqref{eq:M1}, yields a further relation~\cite{Espinosa:2010_EnergyBudget}:
\begin{equation}
\label{eq:speedrel}
\alpha_+ - \frac{1}{3} \left( 1 - \frac{a_-}{a_+} \right) \approx \eta \; \frac{\alpha_+}{\alpha_N} \cdot \frac{1}{2} (v_+ + v_-).
\end{equation}
Eqs.~(\ref{eq:vplus}) and~(\ref{eq:speedrel}) relate $v_+$, $v_-$, and $\alpha_+$, leaving two equations for three unknowns. To resolve this, physical boundary conditions on $\alpha_+$ are applied, depending on the expansion regime.

\subsubsection*{Selection of the bubble expansion regime}

The friction coefficient $\eta$ determines the hydrodynamic regime of the bubble wall expansion. The threshold values $\eta_r$ in Eq. (\ref{eq:etar}) and $\eta_c$ in Eq. (\ref{eq:etac}), separating the two regimes, can be obtained by evaluating Eqs.~(\ref{eq:vplus}) and~(\ref{eq:speedrel}) in the detonation branch under the corresponding limiting conditions:
\begin{itemize}
    \item \textbf{Detonation–runaway transition.}  
    Runaway occurs when the wall velocity approaches the speed of light, $v_w \to 1$. Imposing $v_+ = v_- = 1$ gives
    \begin{equation}
        \eta_r=\alpha_N-\frac{1}{3}\left(1-\frac{a_-}{a_+}\right)\, .
        \label{eq:etar}
    \end{equation}

    \item \textbf{Detonation–deflagration transition.}  
    The transition to subsonic expansion occurs at the Jouguet point, $v_- = c_s$. Imposing this condition gives
    \begin{equation}
        \eta_c = \frac{2}{v_J + c_s}\,\eta_r\, ,
        \label{eq:etac}
    \end{equation}
    where $v_J = \lim_{v_-\rightarrow c_s}v_+$ is given in Eq.~\eqref{eq:vJ}.
\end{itemize}
The bubble expansion regime can therefore be classified as follows:
\begin{itemize}
    \item \textbf{Runaway:} For $\eta \ge \eta_r$, the wall accelerates indefinitely ($v_w \to 1$) and $\alpha_+ = \alpha_N$.
    \item \textbf{Detonation:} For $\eta_c < \eta < \eta_r$, with $v_w = v_+ > c_s$ and $\alpha_+ = \alpha_N$.
    \item \textbf{Deflagration or hybrid:} For $\eta \leq \eta_c$, $\alpha_+ \neq \alpha_N$, with $v_w \leq c_s$. At the boundary $\eta = \eta_c$, one has $v_- = c_s$ and $\alpha_+ = \alpha_N$, and in the hybrid case, we have $v_- = c_s$ but $\alpha_+ \neq \alpha_N$.
\end{itemize}

\subsubsection*{Finding the relevant stationary solution}

Once the regime has been selected, we proceed to solve the hydrodynamics equations in various steps depending on the previous choice, first for the detonation runaway case, we simply need to solve algebraic relations :

\begin{itemize}
    \item {Runaway:} With $v_w = 1$ and $\alpha_+ = \alpha_N$, Eqs.~(\ref{eq:vplus}) and~(\ref{eq:speedrel}) are satisfied trivially. The system~(\ref{eq:edo}) can then be integrated directly to compute $\kappa$.
    \item {Detonation:} For $v_+ = v_w$ and $\alpha_+ = \alpha_N$, Eqs.~(\ref{eq:vplus}) and~(\ref{eq:speedrel}) fully determine $v_-$ and $\kappa$.
\end{itemize}

\noindent In the deflagration and hybrid case $\alpha_+ \neq \alpha_N$, which necessitates instead an iterative approach to determine the proper stationary solution:
\begin{itemize}
\item {Deflagration:} In this case $\alpha_+ \neq \alpha_N$, which requires an iterative approach to determine the proper stationary solution:
\begin{enumerate}
    \item For a given $\alpha_N$ and $\eta$, we scan $\alpha_+$ using a nested sampling algorithm~\cite{dynesty}.
    \item For each trial $\alpha_+$, we solve Eqs.~(\ref{eq:vplus}) and~(\ref{eq:speedrel}) to obtain $v_\pm$ and determine initial conditions at $\xi = \xi_w = v_-$.
    \item The velocity profile $v(\xi)$ is evolved using Eq.~(\ref{eq:edo}) to locate the shock front $\xi_{\rm sh}$ where $v = 0$.
    \item The enthalpy ratio at the shock is~\cite{Espinosa:2010_EnergyBudget},
    \begin{equation}
    \frac{w_{\rm sh}}{w_N} = \frac{9 \xi_{\rm sh}^2 - 1}{3(1 - \xi_{\rm sh}^2)} \;, 
    \end{equation}
    enables the determination of $\alpha_N$ via
    \begin{equation}
    \alpha_N = \alpha_+ \frac{w_+}{w_N}.
    \end{equation}
    \item The iterative process converges when the computed $\alpha_N$ matches the input $\alpha_N$ within a specified tolerance.
\end{enumerate}
\item {Hybrid:} For this final case, the same procedure applies, but with $\xi_w$ treated as a free parameter instead of $\alpha_+$ (since we can obtain the latter directly from Eqs.~(\ref{eq:vplus}) and~(\ref{eq:speedrel}) as $v_- = c_s$ in this regime).
\end{itemize}

Since the whole process starts with the search for steady-state solutions, it cannot yield information on which one of these solutions will be the selected one, in case several of them can be simultaneously obtained. Direct simulations pointed out that runaway walls may occur even for a parameter space for which steady-state solutions can be found~\cite{Krajewski:2024gma}, and it was well-known that several solutions can be found simultaneously with the $\eta$-friction approach~\cite{Espinosa:2010_EnergyBudget}. In this work, we have chosen to select the runaway case when no detonation solutions could be found for large $\alpha_N$, then detonations and finally deflagrations or hybrids (although the last case only occurred in competition to a runaway wall, so that we did not use it in practice). This is an intrinsic uncertainty to the approaches which do not use full hydrodynamic simulations, but we note that it only marginally impacted our predicted reach for the Einstein Telescope, as most solutions there could be found only in a runaway regime.

Finally, we note that adopting the percolation temperature as our reference scale would likely improve the precision of our results. Since $T_P$ depends on the wall velocity, however, hydrodynamic simulations are required to capture this thermal parameter. In an idealised setup, we could incorporate the $v_w$-dependent $T_P$ into our pipeline in the following manner: using the input variables of $\eta$, $\alpha_N$, $T_N$, and $\beta/H_N$ as seed values, we could run the pipeline to obtain an approximate value for $v_w$, after which we could recompute the thermal parameters $\alpha_*$ and $\beta/H_*$ with respect to $T_P$. Note that in such cases, we could include the $v_w$ dependence of the friction. We will leave this for a future work.

\end{appendices}

\bibliographystyle{apsrev4-2}
\bibliography{zotero}

\end{document}